\documentclass[aps,prb,twocolumn,superscriptaddress,floatfix]{revtex4}

\usepackage[vcentermath]{youngtab}
\usepackage{amscd,rotating}
\usepackage{amsmath} 
\usepackage{amssymb}
\usepackage{bm}
\usepackage{color}
\usepackage{graphicx}     
\usepackage{longtable}

\begin{document}

\title{Intermediate coupling fixed point study in the overscreened \\
regime of generalized  multichannel SU($N$) Kondo models}

\author{Damien Bensimon}
\affiliation{Commissariat \`{a} l'Energie Atomique, 
DRFMC /SPSMS, 17 rue des Martyrs, 38054 Grenoble Cedex 9, France}
\affiliation{Max-Planck-Institut f\"ur Festk\"orperforschung,  
Heisenbergstrasse 1, D-70569 Stuttgart, Germany}

\author{Andr\'es Jerez}
\affiliation{Institut Laue Langevin, 6 Rue Jules Horowitz, 
38042 Grenoble Cedex 9, France}
\affiliation{Center for Materials Theory, Rutgers University, 
Piscataway, New Jersey 08854-8019, USA}

\author{Mireille Lavagna}
\altaffiliation{Also at the Centre National de la Recherche Scientifique (CNRS).}
\affiliation{Commissariat \`{a} l'Energie Atomique, 
DRFMC /SPSMS, 17 rue des Martyrs, 38054 Grenoble Cedex 9, France}

\date{\today}


\begin{abstract}
We study a generalized multichannel single-impurity Kondo model, in which the impurity spin is described by a representation of the SU($N$) group which combines bosonic and fermionic degrees of freedom. The impurity spin states are described by Abrikosov pseudofermions, and we make use of a method initiated by Popov and Fedotov which allows a proper handling of the fermionic constraint. The partition function is  derived within a path integral approach. We use renormalization group techniques to calculate the $\beta$ scaling function perturbatively in powers of the Kondo coupling constant, which is justified in the weak coupling limit. The truncated expansion is valid in the overscreened (Nozi\`eres-Blandin) regime, for an arbitrary SU($N$) group and any value of the parameters characterizing the impurity spin representation. The intermediate coupling fixed point is identified. We derive the temperature dependence of various physical quantities at low $T$, controlled by a unique critical exponent, and show that the physics of the system in the overscreened regime governed by the intermediate coupling fixed point is characterized by a non-Fermi liquid behavior. Our results are in accordance with those obtained by other methods, as Bethe ansatz and boundary conformal field theory, in the case of various impurity spin symmetries. We establish in a unified way that the Kondo models in which the impurity spin is described successively by a fundamental, symmetric, antisymmetric and mixed symmetry representation yield all the same low-energy physics in the overscreened regime. Possible generalizations of the analysis we present to the case of arbitrary impurity spin representations of SU($N$) are also discussed.
\end{abstract} 

\maketitle


\section{Introduction}
\label{Intro}

In spite of its twenty-five years of history, the multichannel Kondo model \cite{BlandNoz-1980,Noz-1985} is still a subject of current interest in several fields of study of condensed matter physics. Indeed, this simple impurity model exhibits a non-Fermi liquid (NFL) behavior. Originally introduced to describe metallic systems containing dilute magnetic impurities, its applications have been considerably extended. For instance, it is related to two-level systems. \cite{Zawa-1980,Zawa-1982,ZarVlad-1996} It has been proposed \cite{Zarand-1996} that materials such as $\mathrm{Pb}_{1-x}\mathrm{Ge}_{x}\mathrm{Te}$ and $\mathrm{K}_{1-x}\mathrm{Li}_{x}\mathrm{Cl}$ could be modeled using a multichannel version of the Coqblin-Schrieffer model, \cite{CoqSchri} i.e., a SU($N$)$\times$SU($K$) Kondo model. Here $N$ indicates the number of spin degrees of freedom, and $K$ is the number of channels for the conduction electrons. The multichannel Kondo model is also believed to be relevant to the problem of mesoscopic quantum dots. \cite{Furu-1995,jaz} Several additional possible realizations concern the context of heavy-fermion compounds. Recent experiments in such systems have shown the existence of a quantum phase transition \cite{Sachdev-book} from a magnetically disordered to a long-range magnetic ordered phase, driven by a change in chemical composition, pressure or magnetic field. One can cite the alloy $\mathrm{CeCu}_{5.9}\mathrm{Au}_{0.1}$, \cite{Lohn-1994} the behavior of which in the disordered phase close to the quantum critical point is clearly of NFL-type: the specific heat $C$ depends on $T$ as $C/T\sim -\ln(T/T_{0})$, the magnetic susceptibility as $\chi \sim 1-\alpha \sqrt{T}$, and the $T$-dependent part of the resistivity as $\Delta \rho \sim T$ instead of $C/T \sim \mathrm{Const.}$, $\chi \sim \mathrm{Const.}$ and $\Delta \rho \sim T^{2}$ as predicted by the Fermi liquid theory. \cite{NozPin} Similar types of behavior have been observed in other systems such as $\mathrm{CePd}_{2}\mathrm{Si}_{2}$, \cite{mathur98} and $\mathrm{YbRh}_{2}\mathrm{Si}_{2}$, \cite{trovarelli00} among others (for an extensive survey of the experimental situation concerning heavy-fermion compounds, the reader is refered to the review article of Stewart \cite{Stewart-1994}). The experimentally observed breakdown of the Fermi liquid theory addresses fundamental questions about the possible formation of novel electronic states of matter with new types of elementary excitations resulting from the presence of strong correlations among electrons. \\
\indent
Since it has been proposed the multichannel Kondo model has been studied intensively by a variety of theoretical methods, such as the numerical renormalization group (RG), \cite{Noz-1980,Kusu-1996} Bethe ansatz, \cite{Andrei-1984,TsveWi-1984} conformal field theory, \cite{Frad-1990,Tsve-1990,AfLu-PRL1991,LuAf-PRL1991} and perturbative RG. \cite{Gan-PRL,Gan-thesis} Detailed discusssions concerning the various experimental applications and theoretical aspects of the multichannel Kondo model can be found in the book of Hewson, \cite{Hewson-book} and the review article of Cox and Zawadowski. \cite{CoxZaw} \\
\indent
Recent experiments performed in the heavy-fermion compound $\mathrm{CeCu}_{5.9}\mathrm{Au}_{0.1}$ by inelastic neutron scattering have shown the existence of an anomalous $\omega/T$ scaling law for the dynamical spin suceptibility at the antiferromagnetic wavevector which persists over the entire Brillouin zone. \cite{schroder98,schroder00} Such a result indicates that the spin dynamics are critical not only at large length scales, as studied in the traditional itinerant magnetism picture, \cite{hertz76,millis93,continentino93,moriya95,lp00} but also at atomic length scales. It strongly suggests the presence of locally critical modes beyond the standard spin-fluctuation theory. An approach which has been put forward to describe such a local quantum critical point is based on extended dynamical mean field theory, with encouraging results like the prediction of a scaling law for the dynamical spin suceptibility. \cite{si01} Another theory is based on the generalization of the spin operators symmetry, from SU(2) to SU($N$), to describe the Kondo impurity spin in terms of a mixed fermionic-bosonic representation. \cite{CPT-PRB} The idea behind this mixed formulation is to account for both the extended quasiparticle excitations and the locally critical modes using the fermionic and the bosonic part of the spin, respectively. \\
\indent
In order to fix ideas, let us start with the antiferromagnetic single-channel SU(2) Kondo impurity model. It is well known that within a RG analysis, \cite{Anderson-PMS,Wilson-1974,Wilson-1975,Nozieres-1976} the flow takes the Kondo coupling $J_{\mathrm{K}}$ all the way to strong coupling (i.e., to infinity). An important aspect in the discussion of the breakdown of the Fermi liquid theory is related to the question of the stability of the strong coupling (SC) fixed point. In this case the system flows to a SC fixed point which is stable and the associated behavior of the system is that of a local Fermi liquid. \cite{Noz-LocFL1974,Noz-LocFL1975,Hewson-1980} The situation is rather different when one considers several channels of conduction electrons. In the case of a spin $S$ interacting antiferromagnetically with the spin of conduction electrons belonging to $K$ different channels, Nozi\`eres and Blandin \cite{BlandNoz-1980} have shown that the multichannel Kondo model can lead to two very different situations depending on how $K$ compares to $2S$. Their calculation is based on a second order perturbation theory in the hopping amplitude of the conduction electrons, around the SC fixed point. They analyze their results by deriving an effective coupling between the spin of the composite formed by the impurity dressed by the conduction electrons in the SC limit, and the spin of the conduction electrons on the neighboring sites. On the one hand in the underscreened regime, when $K<2S$, the effective coupling is found to be ferromagnetic and the SC fixed point is stable. In this regime an effective spin, $(S-K/2)$, is formed, resulting from the screening of the impurity spin by the conduction electrons located on the same site. The system, described by the SC fixed point, behaves as a local Fermi liquid. On the other hand in the overscreened regime, when $K>2S$, the effective coupling is found to be antiferromagnetic and hence the SC fixed point is unstable. The instability of the SC fixed point in this regime constitutes a hint for the existence of an intermediate coupling (IC) fixed point, characterized by a finite critical value of the Kondo coupling and leading to a NFL excitation spectrum. \\
\indent
In a previous work \cite{JLB-Kondo03,JLB-Kondo03Short} using a second order perturbation theory similar to the analysis of Nozi\`eres and Blandin, we have considered an antiferromagnetic single-channel SU($N$) Kondo model involving a generalized impurity spin, realized by a combination of $2S$ bosonic and $q$ fermionic degrees of freedom. Preceding studies \cite{CPT-NPB} have shown that the SC fixed point associated with this model includes two regimes, one of them characterized by an instability for a certain class of mixed symmetry impurity spins. More precisely in the large-$N$ limit, \cite{ReadNew-JPC1983} the SC fixed point is stable and describes the low-energy physics of the one conduction channel model when $q < N/2$, but becomes unstable for $q > N/2$. This result suggests that the instability of the SC fixed point can arise from the symmetry properties of the impurity spin, without any introduction of additional degrees of freedom like conduction electron channels, coupling anisotropy or else. We have investigated how the system behaves, in particular when the number of conduction electrons (filling) on the neighboring site of the impurity located at the origin varies.  We have expressed our results in terms of an effective hamiltonian describing the interactions between the impurity dressed by the conduction electrons at the origin (as obtained in the SC limit), and the conduction electrons located on the neighbouring site. By determining the effective couplings, we have shown that the SC instability,  resulting from the change in sign of the effective spin interaction, is related to the change from repulsive to attractive of the effective charge interaction. \\
\indent
In the present work we consider the multichannel version of the model described above, i.e., a SU($N$)$\times$SU($K$) Kondo model, in order to achieve an even richer situa\-tion. In the overscreened regime, the magnetic degrees of freedom of the impurity are completely quenched, and the SC fixed point is unstable. \cite{BlandNoz-1980} We present a method allowing the identification and the characterization of the IC fixed point in the overscreened regime. Our study is based on a perturbative RG method pioneered by Ander\-son, the  ``poor man's scaling'' approach. \cite{Hewson-book,Anderson-PMS} Its basic principle is to reduce the conduction electron band width, $D$, by integrating out the high-energy intermediate states, and letting the low-energy physics properties unchanged. The main difficulty for the description of spin systems is that spin operators obey neither Fermi nor Bose statistics, leading to the absence of a Wick's decomposition theorem in this context. Consequently, conventional perturbative methods such as a Feynman's diagram \- approach are prohibited. The Gaudin theorem \cite{Gaudin-1960} existing for SU(2) generators does not allow the construction of simple diagrammatic techniques directly for spin operators. We tackle this problem by making use of a method initiated by Popov and Fedotov. \cite{PopovFed} In the framework of a fermionic description of the impurity spin states, this powerful approach allows a proper handling of the fermionic constraint. It introduces a set of auxiliary chemical potentials, which eliminates the unphysical Hilbert sub-space without removing the true spin states. The overscreened regime of the Kondo model has been studied intensively by several methods, for various impurity spin representations in both the SU(2) \cite{Gan-PRL,Gan-thesis,FabriZar-1996,Shaw-1998,AffLud-NPB1991} and SU($N$) \cite{jaz,AffLud-1994,Natan-1998,CoRu,pgks} spin groups. Our work considers several of these impurity spin representations as particular limits of the mixed symmetry formulation, enabling direct comparisons with the results obtained by other methods. \\
\indent
This paper is organized as follows. In Section \ref{Sec-KondoMod} the model is introduced, and we present the formalism leading to the derivation of the partition function within a path integral approach. We adopt the Abrikosov pseudofermion formulation to describe the impurity spin states. The handling of the fermionic constraint, performed by making use of the Popov-Fedotov method, is discussed. In Section \ref{Sec-SelfEn} we use a diagrammatic method to derive the conduction electron self-energy perturbatively in powers of the Kondo coupling constant, i.e., in the weak coupling limit. The imaginary part of the conduction electron self-energy is proportional to the electronic scattering rate, which is invariant under RG transformation as any physical quantity. This invariance is exploited in Section \ref{Sec-PRenGr}, to compute the $\beta$ scaling function by making use of perturbative RG arguments, including the  Callan-Symanzik equation. This leads to the identification of the IC fixed point in the overscreened regime. The temperature dependence of various physical quantities at low $T$, controlled by a unique critical exponent, is derived. The NFL nature of the physics governed by the IC fixed point is also discussed, and our results are compared with those of previous works. Section \ref{Conclu} gives a summary of the method and main results, and is devoted to conclusions. \\
\indent
The appendices contain the technical details of the calculations. In Appendix \ref{Appen-SUNgener} we give some properties followed by the SU($N$) group generators. Appendix \ref{Appen-Popov} briefly explains the principle of the method pioneered by Popov and Fedotov. This method requires a particular derivation of the conduction electron and impurity spin pseudofermion free propagators, which is performed in Appendix \ref{Appen-FreeProp}. Appendix \ref{Appen-SpinFact} contains the calculation of the first order term spin factors showing up in the conduction electron self-energy expansion. In Appendix \ref{Appen-ThirOrd} we show exhaustively how to compute the contribution to the electronic self-energy for a given example of diagram. The other diagrams can then be calculated in a similar way.


\section{Generalized Kondo Model and formalism}
\label{Sec-KondoMod}

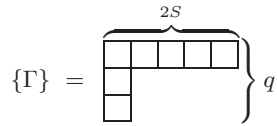
\begin{figure}[t]
\begin{eqnarray*} 
\{ \Gamma \}~=~ \overbrace{\yng(5,1,1)}^{2S} 
\begin{picture}(40,40)
\put(-9.0,0.1){$\left. \begin{array}{l} \\ \\ \\ \end{array} \right\} q$ }
\end{picture}
\end{eqnarray*}
\caption {Young Tableau description of the L-shaped irreducible representation $\{ \Gamma \}$ associated with the mixed symmetry impurity spin, realized by a combination of bosonic and fermionic degrees of freedom.}
\label{LShape-Fig} 
\end{figure}

In this Section we present the model, and the methods used to derive the partition function which is obtained in the functional integration formalism. In the language of Young tableaus, the impurity spin is described by a L-shaped representation, as shown in Fig.\! \ref{LShape-Fig}. The bosonic degrees of freedom correspond to the horizontal line (first row) containing $2S$ boxes, while the fermionic ones are included in the $q$ boxes of the vertical line (first column). 

\subsection{Generalized multichannel SU($N$) single-impurity Kondo model}
\label{Sub-Kondo}

 The Hamiltonian describing the Kondo model we consider is
\begin{eqnarray}
\mathcal{H}_{\mathrm{K}} = \mathcal{H}_{0} + \mathcal{V}_{\mathrm{K}}~,  
\label{HK-def}
\end{eqnarray}

\noindent where $\mathcal{H}_{0}$ is the free electronic part and $\mathcal{V}_{\mathrm{K}}$ gives the interaction between the generalized impurity spin and the conduction electrons. A SU($N$) (for $N \geq 2$) spin degeneracy is required for both the impurity and the conduction electron spins, and we suppose that there exists $K$ ($\geq 1$) electronic conduction channel(s)
\begin{eqnarray}
\mathcal{H}_{0} = \sum_{\bm{k}} \sum_{\sigma=1}^{N}  \sum_{\lambda=1}^{K} 
\varepsilon _{\bm{k}} c_{\bm{k},\sigma,\lambda}^{\dagger} c_{\bm{k},\sigma,\lambda}~,
\label{H0-def}
\end{eqnarray}
where $c_{\bm{k},\sigma,\lambda}^{\dagger}$ is the creation operator of a conduction electron with momentum $\bm{k}$ and energy $\varepsilon _{\bm{k}}$, $\sigma$ and  $\lambda$ are the spin and channel indices, respectively. The second term  
\begin{eqnarray}
\mathcal{V}_{\mathrm{K}} = J_{\mathrm{K}} \vec{\bm{S}}_{i} \cdot  \vec{\bm{S}}_{e}(0)~,
\label{VK-def}
\end{eqnarray}
corresponds to the antiferromagnetic Kondo interaction $J_{\mathrm{K}}$ ($\geq 0$) between the impurity spin $\vec{\bm{S}}_{i}$ located at the origin (site 0) and the spin of the conduction electrons $\vec{\bm{S}}_{e}(0)$ located at the same site. \cite{Kondo-1964}

The impurity spin states corresponding to the irreducible representation $\{ \Gamma \}$ can be expressed in terms of Abrikosov pseudofermions. \cite{Abri1965} We denote the dimension of $\{ \Gamma \}$ by $d_{ \{ \Gamma \} }$. Its general expression is given in Appendix \ref{Appen-SUNgener}. Each impurity spin state is described by a fermion. Therefore we have $d_{ \{ \Gamma \} }$ species of pseudofermions located at the origin which are respectively created and destroyed by the following operators
\begin{eqnarray}
f_{\gamma}^{\dagger} \ , \ f_{\gamma} \ , \ 1 \leq \gamma \leq d_{ \{ \Gamma \} }~.
\label{Def-fermi}
\end{eqnarray}
The number of pseudofermions at the impurity site is subject to the standard constraint
\begin{eqnarray}
\hat{n}_{f} = \sum_{\gamma=1}^{d_{ \{ \Gamma \} } } f_{\gamma}^{\dagger}  f_{\gamma} = 1~.
\label{Cons-fermi}
\end{eqnarray}
\noindent According to the structure of the SU($N$) Lie group, \cite{Licht1970} the impurity spin operator owns ($N^{2}-1$) components
\begin{eqnarray}
\vec{\bm{S}}_{i} =  \Big[ \bm{S}_{i}^{A} \Big]_{1\leq A \leq N^{2}-1}~, \nonumber
\end{eqnarray}
with
\begin{eqnarray}
\bm{S}_{i}^{A} =  \sum_{\gamma,\gamma^{'}=1}^{d_{ \{ \Gamma \} } }
f_{\gamma^{'}}^{\dagger} \bm{\Gamma}_{\gamma^{'},\gamma}^{A} f_{\gamma}~,
\label{Imp-compo}
\end{eqnarray}
where $\bm{\Gamma}^{A}$ corresponds to the ($N^{2}-1$) generators of the SU($N$) spin group.
A crucial technical aspect has to be pointed out here. Due to the shape of the generalized impurity spin that we consider which is not in the fundamental SU($N$) representation (i.e., a single box representation in the language of Young tableaus), and to the Abrikosov pseudofermion formalism that we adopt, the $\big\{ \tilde{\bm{\Gamma}}^{A} \big\}_{1\leq A \leq N^{2}-1}$ SU($N$) generators are not expanded as ($N \times N$) matrices as usual but as $\big( d_{ \{ \Gamma \} } \times d_{ \{ \Gamma \} }\big)$ ones. Such a situation already occurs in the SU(2) group if instead of considering a spin 1/2 corresponding to the SU(2) fundamental representation, we work with a spin 1, 3/2, ... \cite{GreiMull} The properties of the $\big\{ \tilde{\bm{\Gamma}}^{A} \big\}$ matrices which we will need are given in Appendix \ref{Appen-SUNgener}. It is to be noted that this way of writing the impurity spin states, although equivalent, is completely different from the method that we used in our previous work. \cite{JLB-Kondo03,JLB-Kondo03Short} In the latter case, the impurity spin states were obtained as a combination of ($2S+q-1$) particles, each box of the L-shaped representation being described by either one boson or one fermion, and the SU($N$) generators we had to consider corresponded to the fundamental representation. However the approach we now consider in terms of pure pseudofermions is much more appropriate to the functional integration formalism we develop. 

\begin{figure}[t]
\begin{eqnarray*}
\{ \sigma \}~=~ \yng(1)
\end{eqnarray*}
\caption{Young Tableau description of the fundamental (irreducible) representation $\{ \sigma \}$ associated with each conduction electron.}
\label{FundaRep-Fig} 
\end{figure}

As shown in Fig.\! \ref{FundaRep-Fig}, each conduction electron is described by the fundamental representation of SU($N$). The spin of the conduction electrons at the origin $\vec{\bm{S}}_{e}(0)$ is given by
\begin{eqnarray}
\vec{\bm{S}}_{e}(0) = \Big[ \bm{S}_{e}^{A}(0) \Big]_{1\leq A \leq N^{2}-1}, \nonumber
\end{eqnarray}
with the $(N^{2}-1)$ spin components given by
\begin{eqnarray}
\bm{S}_{e}^{A}(0) = \sum_{\sigma,\sigma^{'}=1}^{N}  \sum_{\lambda=1}^{K} 
c_{\sigma^{'},\lambda}^{\dagger}(0) \bm{\sigma}_{\sigma^{'},\sigma}^{A} c_{\sigma,\lambda}(0)~. \nonumber
\end{eqnarray}
where $\bm{\sigma}^{A}$ corresponds to the ($N^{2}-1$) generators of the SU($N$) spin group.
In the previous expression $c_{\sigma^{'},\lambda}^{\dagger}(0)$ is the creation operator of a conduction electron of spin color $\sigma^{'}$ in the channel $\lambda$ at the impurity site, which can be rewritten in the momentum space as
\begin{eqnarray}
c_{\sigma^{'},\lambda}^{\dagger}(0) = \frac{1}{\sqrt{\mathcal{N}_{s}}} \sum_{\bm{k}} c_{\bm{k},\sigma^{'},\lambda}^{\dagger}~,
\label{Def-ElOp}
\end{eqnarray}
where $\mathcal{N}_{s}$ is the number of sites of the lattice. Using Eq.\! (\ref{Def-ElOp}), the different components of the conduction electron spin can be expressed as
\begin{eqnarray}
\bm{S}_{e}^{A}(0) = \frac{1}{\mathcal{N}_{s}} \sum_{\bm{k},\bm{k}^{'}} \sum_{\sigma,\sigma^{'}=1}^{N}  \sum_{\lambda=1}^{K} 
c_{\bm{k}^{'},\sigma^{'},\lambda}^{\dagger} \bm{\sigma}_{\sigma^{'},\sigma}^{A} c_{\bm{k},\sigma,\lambda}~. 
\label{El-compo}
\end{eqnarray}
The SU($N$) generators for the conduction electron spin correspond to the fundamental representation, therefore the corresponding traceless hermitian matrices $\big\{ \tilde{\bm{\sigma}}^{A} \big\}_{1\leq A \leq N^{2}-1}$ have ($N \times N$) components and satisfy the following normalization equation
\begin{eqnarray}
\mathrm{Tr}\big[ \tilde{\bm{\sigma}}^{A} \tilde{\bm{\sigma}}^{B} \big] = \frac{1}{2} \delta_{AB}~. \nonumber
\end{eqnarray}
\noindent In the well known SU(2) and SU(3) groups, $\tilde{\bm{\sigma}}^{A} = \tilde{\bm{\tau}}^{A}/2$ where $\big\{ \tilde{\bm{\tau}}^{A} \big\}$ are the Pauli and Gell-Mann matrices, respectively.
 
By making use of Eqs.\! (\ref{H0-def}), (\ref{VK-def}), (\ref{Imp-compo}) and (\ref{El-compo}) we obtain the Hamiltonian related to the multichannel SU($N$) $\times$ SU($K$) single-impurity Kondo model for a L-shaped Young tableau representation of the impurity spin 
\begin{eqnarray}
\mathcal{H}_{\mathrm{K}} = & & \sum_{\bm{k}} \sum_{\sigma=1}^{N}  \sum_{\lambda=1}^{K} 
\varepsilon _{\bm{k}} c_{\bm{k},\sigma,\lambda}^{\dagger} c_{\bm{k},\sigma,\lambda} \nonumber \\
&+& \frac{J_{\mathrm{K}}}{\mathcal{N}_{s}} \sum_{A=1}^{N^{2}-1}
\Big( \sum_{\bm{k},\bm{k}^{'}} \sum_{\sigma,\sigma^{'}=1}^{N}  \sum_{\lambda=1}^{K} 
c_{\bm{k}^{'},\sigma^{'},\lambda}^{\dagger} \bm{\sigma}_{\sigma^{'},\sigma}^{A} c_{\bm{k},\sigma,\lambda} \Big) \nonumber \\
& & \hspace{14mm} \times \Big( \sum_{\gamma,\gamma^{'}=1}^{d_{ \{ \Gamma \} } }
f_{\gamma^{'}}^{\dagger} \bm{\Gamma}_{\gamma^{'},\gamma}^{A} f_{\gamma} \Big)~.
\label{HK-full}
\end{eqnarray}


\subsection{Functional integration method - Handling of the fermionic constraint}
\label{Sub-FuInt}

The partition function corresponding to $\mathcal{H}_{\mathrm{K}}$ (\ref{HK-full}) is
\begin{eqnarray}
\mathcal{Z} = \mathrm{Tr}_{\mbox{\scriptsize{spin}}} \mathrm{exp} \big[- \beta \mathcal{H}_{\mathrm{K}} \big]~,
\label{Tr-PartFu}
\end{eqnarray}
with $\beta = 1/k_{\mathrm{B}}T$. The difficulty in computing $\mathcal{Z}$ comes from the constraint expressed in Eq.\! (\ref{Cons-fermi}) which requires to perform the trace $\mathrm{Tr}_{\mbox{\scriptsize{spin}}}$ over the physical states involving one pseudofermion at the impurity site only. However the Pauli principle allows the number of pseudofermions at the origin to vary between 0 and $d_{ \{ \Gamma \} }$. More explicitly, the representation of the impurity spin with pseudofermion operators enlarges the dimensionality of the Hilbert space in which these fields are acting, compared with the physical space appropriate for the true spin operators. The unphysical states displayed in the pseudofermion formulation are then eliminated with the aid of the constraint. Therefore the evaluation of the trace in Eq.\! (\ref{Tr-PartFu}) can not be simply performed using standard perturbation approaches for instance but requires the use of further techniques.

In order to settle this problem, we adopt a method pioneered by Popov and Fedotov.  \cite{PopovFed} As explained in Appendix \ref{Appen-Popov}, one can introduce a set of auxiliary (imaginary) chemical potentials  $\{ \mathrm{i} \mu_{n} \}_{1\leq n \leq  d_{ \{ \Gamma \} }}$ governed by a discrete distribution function $P\{ \mu \}$
\begin{eqnarray}
& &P\{ \mu \} = \frac{1}{d_{ \{ \Gamma \} }} \sum_{n=1}^{d_{ \{ \Gamma \} } }  
\mathrm{exp} \left[\mathrm{i} \beta \mu_{n} \right]
\delta \big( \mu+\mathrm{i} \mu_{n} \big) \nonumber \ , \\
& &\mathrm{i} \mu_{n} = \frac{\mathrm{i} \pi}{\beta d_{ \{ \Gamma \} }} (2n-1)~.
\label{Def-Pmu}
\end{eqnarray}
A crucial property arising from Eq.\! (\ref{Def-Pmu}) is that the imaginary chemical potentials depend on temperature. The partition function can be rewritten as
\begin{eqnarray}
\mathcal{Z} = \int \mathcal{D}\mu \ P\{ \mu \} \ \mathrm{Tr}_{\mbox{\scriptsize{Fock}}} \mathrm{exp} 
\Big[- \beta \big( \mathcal{H}_{\mathrm{K}}- \mu \hat{n}_{f} \big) \Big]~,
 \label{Pmu-PartFu}
\end{eqnarray}
where the trace $\mathrm{Tr}_{\mbox{\scriptsize{Fock}}}$ is performed over the entire Fock space corresponding to the fermionic impurity spin operators without any constraint, namely including any number of pseudofer\-mion(s) allowed by the Pauli principle. The contributions of the unphysical states ($n_{f} \neq 1$) to the partition function automatically cancel out one with other by making use of the imaginary chemical potentials. With Eqs.\! (\ref{Def-Pmu}) and (\ref{Pmu-PartFu}) we get
\begin{eqnarray}
& &\mathcal{Z} = \frac{1}{d_{ \{ \Gamma \} }} \sum_{n=1}^{d_{ \{ \Gamma \} } }  \mathrm{exp} \left[\mathrm{i} \beta \mu_{n} \right]  \\
& &\hspace{22mm} \times \mathrm{Tr}_{\mbox{\scriptsize{Fock}}} \mathrm{exp} 
\Big[- \beta \big( \mathcal{H}_{\mathrm{K}} + \mathrm{i}\mu_{n} \hat{n}_{f} \big) \Big]~. \nonumber
\label{Sumu-PartFu}
\end{eqnarray}
In the latter equation, in contrast with Eq.\! (\ref{Tr-PartFu}) of $\mathcal{Z}$, the trace showing up in the calculation has to be performed in the grand canonical ensemble, which allows to use standard techniques, namely the functional integration, Green's function method and diagrammatic expansion. In a path integral formalism at finite temperature, \cite{NegOr,Popov-book} each Fock trace is given by
\begin{eqnarray}
& &\mathrm{Tr}_{\mbox{\scriptsize{Fock}}} \mathrm{exp} 
\Big[- \beta \big( \mathcal{H}_{\mathrm{K}} + \mathrm{i}\mu_{n} \hat{n}_{f} \big) \Big] \\
&=& \int \mathcal{D}\bar{c}^{*} \mathcal{D}\bar{c} \mathcal{D}\bar{f}^{*} \mathcal{D}\bar{f}
 \mathrm{exp} \Bigg[- \int_{0}^{\beta}d\tau \big\{ \mathcal{L}_{n}(\tau) + 
\mathcal{H}_{\mathrm{K}}(\tau) \big\} \Bigg]~, \nonumber
\label{Tr-Fock}
\end{eqnarray}
where $\bar{c}^{*}$, $\bar{c}$, $\bar{f}^{*}$, $\bar{f}$ are Grassmann variables associated with the fermionic operators $c^{\dagger}$, $c$, $f^{\dagger}$, $f$, respectively, and $\mathcal{L}_{n}(\tau)$ is the Lagrangian of free particles expressed as 
\begin{eqnarray}
& &\mathcal{L}_{n}(\tau) = \sum_{\bm{k}} \sum_{\sigma=1}^{N}  \sum_{\lambda=1}^{K} 
\bar{c}_{\bm{k},\sigma,\lambda}^{*}(\tau) \partial_{\tau} \bar{c}_{\bm{k},\sigma,\lambda}(\tau) \nonumber \\
& & \hspace{14mm} + \sum_{\gamma=1}^{d_{ \{ \Gamma \} } } \bar{f}_{\gamma}^{*}(\tau) \big( \partial_{\tau} + \mathrm{i}\mu_{n} \big) \bar{f}_{\gamma}(\tau)~.
\label{Free-Lagrang}
\end{eqnarray}
One can notice from Eq.\! (\ref{Free-Lagrang}) that a direct consequence of the introduction of the auxiliary chemical potentials is a shift of the Matsubara frequencies related to the pseudo\-fermions by a value of ($-\mathrm{i}\mu_{n}$).

Using Eqs.\! (\ref{Sumu-PartFu}) and (\ref{Tr-Fock}), we finally obtain the expression of the partition function
\begin{eqnarray}
\label{LnHK-PartFu}
\mathcal{Z} &=& \frac{1}{d_{ \{ \Gamma \} }} \sum_{n=1}^{d_{ \{ \Gamma \} } } \mathrm{exp} \left[\mathrm{i} \beta \mu_{n} \right]   
\int \mathcal{D}\bar{c}^{*} \mathcal{D}\bar{c} \mathcal{D}\bar{f}^{*} \mathcal{D}\bar{f} \\
& &\hspace{15mm} \times \ \mathrm{exp} \Bigg[- \int_{0}^{\beta}d\tau 
\big\{ \mathcal{L}_{n}(\tau) + \mathcal{H}_{\mathrm{K}}(\tau) \big\} \Bigg]~. \nonumber
\end{eqnarray}


\section{Calculation of the conduction electron self-energy}
\label{Sec-SelfEn}
In this Section, we expand the conduction electron self-energy perturbatively in powers of the Kondo coupling constant $J_{\mathrm{K}}$. This means that we consider the weak coupling limit. We closely follow the method developed by Gan \textit{et al.} \cite{Gan-PRL,Gan-thesis} in the special case of the SU(2) $\times$ SU($K$) Kondo model, involving an impurity of spin 1/2. We use diagrammatic techniques. \cite{Mahan,Mattuck} 


\subsection{Construction of the electronic self-energy diagrams}
\label{Sub-ConstDiag}

Here we summarize the rules allowing to derive the self-energy of the conduction electron, using a diagrammatic perturbative expansion of the partition function in $J_{\mathrm{K}}$. A detailed method to build the diagrams associated with the electronic self-energy in the framework of the Kondo model is given by Silverstein and Duke, \cite{Silv-Du} following  Abrikosov. \cite{Abri1965} Although the final results will be given at $T=0$, we work in the space of (fermionic) Matsubara frequencies for technical simplicity. The rules are the following 

(i) In order to get the contribution at the $m$th order in the Kondo interaction $J_{\mathrm{K}}$, i.e., $\mathcal{O} \big[(J_{\mathrm{K}})^{m} \big]$, we draw all the topologically distinct diagrams linking $m$ vertices.

(ii) Each conduction electron propagator (\ref{FreePropEl}) is represented by a solid line  \\
\begin{picture}(40,25)
\put(100.1,7.0){$\bm{k}_{\bm{1}}, \mathrm{i} \omega_{m_{1}^{e}}, \sigma_{1}$}
\put(90.1,0.2){\vector(1,0){40}}
\put(130.1,0.2){\line(1,0){35} \ .}
\end{picture}
\\

(iii) Each impurity spin pseudofermion propagator (\ref{FreePropFermi}) is drawn as a dashed line \\
\begin{picture}(40,25)
\put(110.1,10.0){$\mathrm{i} \omega_{m_{1}^{i}}, \gamma_{1}$}
\put(90.1,0.2){- - - - - - } 
\put(130.1,2.7){\vector(1,0){0.5}}   
\put(132.1,0.2){- - - - - \ .}
\end{picture}
\\

(iv) Each vertex \\
\begin{picture}(40,50)
\put(26.1,26.0){$\bm{k}_{\bm{1}}, \mathrm{i} \omega_{m_{1}^{e}}, \sigma_{1}$}
\put(84.1,26.0){\vector(2,-1){30}}
\put(110.1,13.0){\line(2,-1){20}}
\put(130.1,3.0){\vector(2,1){25}}
\put(150.1,13.0){\line(2,1){30}}
\put(182.1,26.0){$\bm{k}_{\bm{2}}, \mathrm{i} \omega_{m_{2}^{e}}, \sigma_{2}$}
\put(130.1,3.0){\circle*{4.9}}
\put(16.1,0.5){$\mathrm{i} \omega_{m_{1}^{i}}, \gamma_{1}$}
\put(55.1,0.2){- - - - - - } 
\put(95.1,2.7){\vector(1,0){0.5}}   
\put(95.1,0.2){- - - - -}
\put(130.1,0.2){- - - - - - } 
\put(170.1,2.7){\vector(1,0){0.5}}   
\put(170.1,0.2){- - - - -}
\put(205.1,0.5){$\mathrm{i} \omega_{m_{2}^{i}}, \gamma_{2}$ }
\end{picture}
\\ \\
contributes by a factor

\begin{eqnarray}
- \frac{\displaystyle J_{\mathrm{K}}}{\displaystyle \beta \mathcal{N}_{s}} 
\delta_{m_{1}^{e}+m_{1}^{i},m_{2}^{e}+m_{2}^{i}}
\sum\limits_{A=1}^{N^{2}-1}
\bm{\sigma}_{\sigma_{2},\sigma_{1}}^{A}  \bm{\Gamma}_{\gamma_{2},\gamma_{1}}^{A}~.
\label{VertexContrib}
\end{eqnarray}

(v) The independent internal momentums, frequencies and spin indices have to be summed over.

(vi)  Each conduction electron loop gives a factor $(- K)$.

(vii) Each pseudofermion loop contributes by a factor $(-1)$.

(viii) Each diagram $\mathcal{O}\big[(J_{\mathrm{K}})^{m}\big]$ exhibits a combinatorial symmetry factor $f_{sym}$. This factor results from the non-compensation between the number of permutations of the vertex and the contribution (1/$m!$) due to Wick's theorem. \cite{ItzikZub} We show it explicitly in Figs.\! \ref{SelfEnDiags-Fig2}, \ref{SelfEnDiags-Fig3} and \ref{SelfEnDiags-Fig4}, when its value is not equal to 1.


\subsection{Contributions of the leading order terms to the conduction electron self-energy}
\label{Sub-CalcSelfEn}

\begin{figure}[t]
\includegraphics[width=6.0cm]{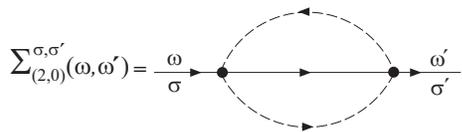}
\caption {Feynman diagram for the conduction electron self-energy at the second order in a perturbation theory in $J_{\mathrm{K}}$.} 
\label{SelfEnDiags-Fig2}
\end{figure}

\begin{figure}[t]
\includegraphics[width=8.6cm]{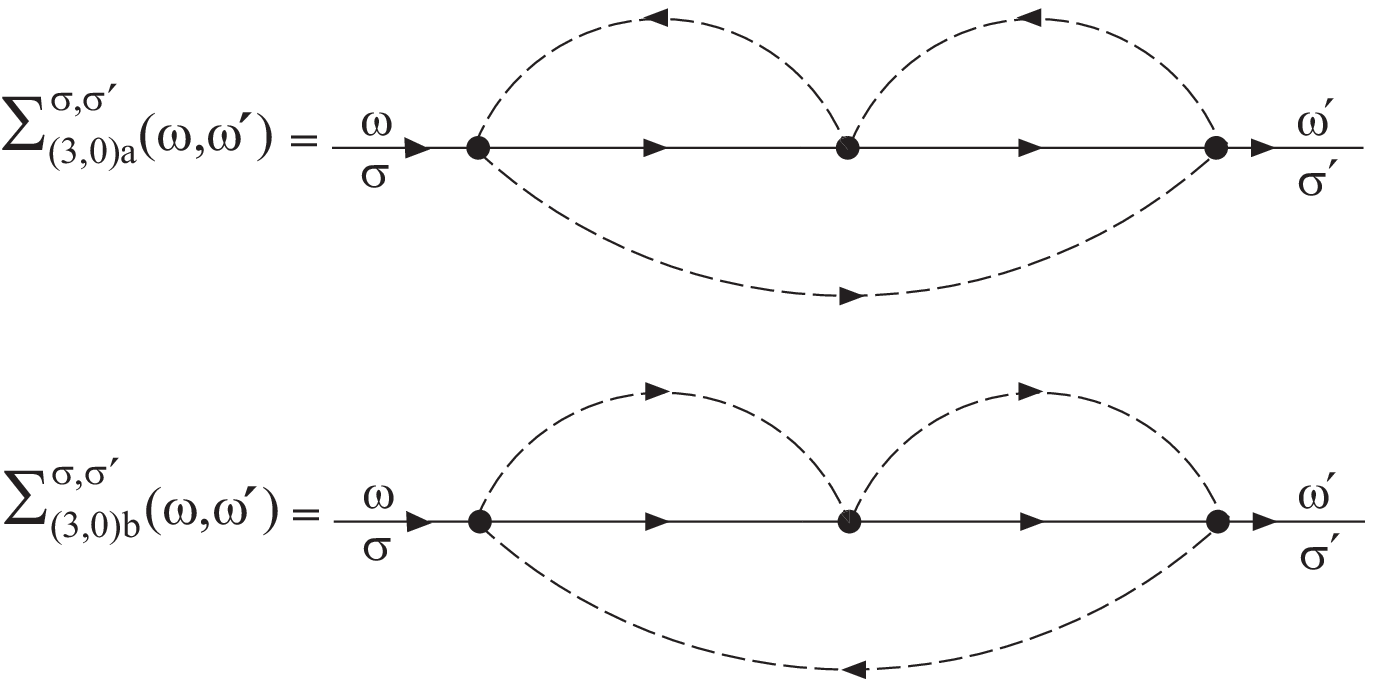}
\caption {Feynman diagrams for the conduction electron self-energy at the third order in a perturbation theory in $J_{\mathrm{K}}$.} 
\label{SelfEnDiags-Fig3}
\end{figure}

\begin{figure}[b]
\includegraphics[width=6.3cm]{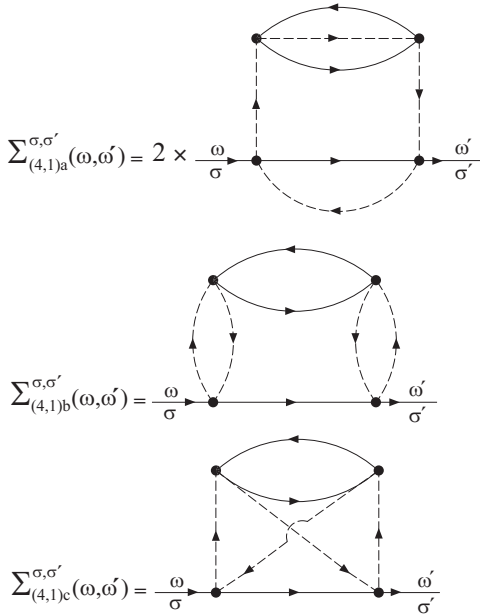}
\caption {Feynman diagrams for the conduction electron self-energy at the fourth order in a perturbation theory in $J_{\mathrm{K}}$.} 
\label{SelfEnDiags-Fig4}
\end{figure}

In Figs.\! \ref{SelfEnDiags-Fig2}, \ref{SelfEnDiags-Fig3}, and \ref{SelfEnDiags-Fig4} we show the diagrams which will allow us to compute the $\beta$ scaling function perturbatively. As discussed in Section \ref{Sec-PRenGr}, these digrams are sufficient to derive the critical exponent of the model. The two subscripts in each term of the self-energy expansion indicate the corresponding powers of $J_{\mathrm{K}}$ and $K$, respectively. Each diagram is factorized along two main terms. The first one is the spin factor, it results from the summation over the internal spin degrees of freedom. In Appendix \ref{Appen-SpinFact} we present the method of calculation of these spin factors for all the diagrams we consider. It requires some knowledge about the properties (detailed in Appendix \ref{Appen-SUNgener}) of the SU($N$) generators $\big\{ \tilde{\bm{\Gamma}} \big\}$  and $\big\{ \tilde{\bm{\sigma}} \big\}$ associated with the impurity and the electronic spins, respectively. The second quantity is obtained by summing over the internal momentum and frequency degrees of freedom. Due to the presence of the imaginary chemical potentials $\{ \mathrm{i} \mu_{n} \}_{1\leq n \leq  d_{ \{ \Gamma \} }}$, this part has to be computed carefully. In particular, the free propagators associated with the conduction electron and impurity spins have to be redefined, which is done in Appendix \ref{Appen-FreeProp}. In Appendix \ref{Appen-ThirOrd}, we detail the calculation of the third order diagram $\mbox{\large{$\Sigma$}}_{(3,0)\mbox{\scriptsize{a}}}^{\sigma, \sigma^{'}}$ displayed in Fig.\! \ref{SelfEnDiags-Fig3}. The other diagrams can be computed in the same way.

The single-impurity Kondo model can be used for the study of real systems of dilute (low concentration) impurities. For that purpose, we adopt the random impurity averaging, \cite{Abri1965} the principle of which is the following. A number $N_{i}$ of impurity spins are randomly distributed in the system. We only keep the linear contribution in the impurity density $n_{i} = N_{i} / \mathcal{N}_{s}$. Since the interactions between impurities are completely neglected, we recover a single-impurity Kondo model.
\indent To summarize, each of the contributions $\mbox{\large{$\Sigma$}}_{(j,l)\alpha}^{\sigma, \sigma^{'}}$ to the conduction electron self-energy corresponding to a diagram of Figs.\! \ref{SelfEnDiags-Fig2}, \ref{SelfEnDiags-Fig3}, or \ref{SelfEnDiags-Fig4} can be written as a factorized expression as follows
\begin{eqnarray}
\mbox{\large{$\Sigma$}}_{(j,l)\alpha}^{\sigma, \sigma^{'}} &=& (-1)^{b_{pf}} \cdot (- K)^{l} \cdot (N_{i}) \cdot  f_{sym} \nonumber \\
& &\times \ \mbox{Spin}\Big[ \mbox{\large{$\Sigma$}}_{(j,l)\alpha}^{\sigma, \sigma^{'}} \Big] \cdot F_{(j,l)\alpha}~,
\label{Decompo-SelfEn}
\end{eqnarray}
where $b_{pf}$ and $l$ are the numbers of pseudofermion and conduction electron loop(s), respectively, $N_{i}$ is the number of impurities in the system, $f_{sym}$ the combinatorial symmetry factor of the diagram, $\mbox{Spin}\Big[ \mbox{\large{$\Sigma$}}_{(j,l)\alpha}^{\sigma, \sigma^{'}} \Big]$ the spin term corresponding to the contribution of the spin operators which involves the summation over the internal spin degrees of freedom, and $F_{(j,l)\alpha}$ the factor resulting from the contribution of the fermionic propagators which is obtained by performing the summations over the internal momentum and Matsubara frequencies. \\
\indent We shall exclusively work at $T = 0$, allowing us to calculate the integrals in energy analytically as well as the summations over imaginary chemical potentials included in the impurity spin free propagator (\ref{FreePropFermi}). We follow the conventional cut-off scheme: a cut-off in energy is introduced for the conduction electron band, i.e., $-D \leq \varepsilon_{\bm{k}} \leq D$, with a density of states $\rho$ per spin and per channel assumed to be constant.
\vfill\eject
\begin{widetext}
At zero-temperature after analytical continuation to the real axis, we obtain the contributions of the first perturbative order terms to the imaginary part of the electronic self-energy
\begin{eqnarray}
\label{SelfEn-2Ord}
& &\mbox{Im} \Big[\mbox{\large{$\Sigma$}}_{(2,0)}^{\sigma, \sigma^{'}} \big(\omega, \omega^{'}\big) \Big]  
= - \Bigg( \frac{(N+2S-1)!}{(2S)! (q-1)! (N-q)!} \cdot \frac{2S}{4 N^{2}}
\Big[    2S \big(N - 1 \big) - q \big(N + 1 \big) +  \big(N^{2}+1 \big)\Big] \Bigg)  \\
& &\hspace{35mm} \times \big( J_{\mathrm{K}} \big)^{2}\rho  \cdot n_{i}  \cdot
\left(\frac{d_{ \{ \Gamma \} }-1}{\big[d_{ \{ \Gamma \} }\big]^{2} }\right) 
\cdot \mbox{Im} \Bigg\{ \int_{-D}^{+D} d \varepsilon_{1} \frac{1}{(\omega -\varepsilon_{1} + \mathrm{i} 0^{+})}  \Bigg\}
\cdot \delta_{\sigma, \sigma^{'}} \cdot \delta \big(\omega-\omega^{'}\big) \ , \nonumber \\
\label{SelfEn-3Ord}
& &\mbox{Im} \Big[\mbox{\large{$\Sigma$}}_{(3,0)\mbox{\scriptsize{a}}}^{\sigma, \sigma^{'}} \big(\omega, \omega^{'}\big) \Big] +
\mbox{Im} \Big[\mbox{\large{$\Sigma$}}_{(3,0)\mbox{\scriptsize{b}}}^{\sigma, \sigma^{'}} \big(\omega, \omega^{'}\big) \Big]  \\
&=& - \Bigg( \frac{(N+2S-1)!}{(2S)! (q-1)! (N-q)!} \cdot \frac{2S}{4 N}
\Big[    2S \big(N - 1 \big) - q \big(N + 1 \big) +  \big(N^{2}+1 \big)\Big] \Bigg) \nonumber \\
& &\hspace{4mm} \times \big( J_{\mathrm{K}} \big)^{3} \rho^{2} \cdot n_{i}  \cdot \left(\frac{d_{ \{ \Gamma \} }-1}{\big[d_{ \{ \Gamma \} }\big]^{2} }\right)
\cdot \mbox{Im} \Bigg\{ \int_{-D}^{+D} d \varepsilon_{1} d \varepsilon_{2} \
\frac{n_{\mathrm{FD}}(\varepsilon_{2})}
{(\omega -\varepsilon_{1} + \mathrm{i} 0^{+})(\varepsilon_{1} - \varepsilon_{2}) } \Bigg\}
\cdot \delta_{\sigma, \sigma^{'}} \cdot \delta \big(\omega-\omega^{'}\big)~, \nonumber \\
\label{SelfEn-4Ord}
& & \mbox{Im} \Big[ \mbox{\large{$\Sigma$}}_{(4,1)\mbox{\scriptsize{a}}}^{\sigma, \sigma^{'}}  \big(\omega, \omega^{'}\big) \Big] +
 \mbox{Im} \Big[ \mbox{\large{$\Sigma$}}_{(4,1)\mbox{\scriptsize{b}}}^{\sigma, \sigma^{'}}  \big(\omega, \omega^{'}\big) \Big] +
 \mbox{Im} \Big[ \mbox{\large{$\Sigma$}}_{(4,1)\mbox{\scriptsize{c}}}^{\sigma, \sigma^{'}}  \big(\omega, \omega^{'}\big) \Big] \\
&=& \Bigg( \frac{(N+2S-1)!}{(2S)! (q-1)! (N-q)!} \cdot \frac{2S}{8 N}
\Big[    2S \big(N - 1 \big) - q \big(N + 1 \big) +  \big(N^{2}+1 \big)\Big] \Bigg) 
\cdot \big( J_{\mathrm{K}} \big)^{4}\rho^{3}  \cdot n_{i} \cdot K 
\cdot \left(\frac{d_{ \{ \Gamma \} }-1}{\big[d_{ \{ \Gamma \} }\big]^{2} }\right) \nonumber \\
& &\hspace{2mm} \times \ \mbox{Im} \Bigg\{ \int_{-D}^{+D} d \varepsilon_{1} d \varepsilon_{2} d \varepsilon_{3} \
\frac{n_{\mathrm{FD}}(-\varepsilon_{1}) n_{\mathrm{FD}}(\varepsilon_{2}) n_{\mathrm{FD}}(-\varepsilon_{3})
+n_{\mathrm{FD}}(\varepsilon_{1}) n_{\mathrm{FD}}(-\varepsilon_{2}) n_{\mathrm{FD}}(\varepsilon_{3})  }
{(\omega -\varepsilon_{1} + \mathrm{i} 0^{+})
(\omega -\varepsilon_{1} +\varepsilon_{2} - \varepsilon_{3} + \mathrm{i} 0^{+})(\varepsilon_{2} - \varepsilon_{3}) } \Bigg\}
\cdot \delta_{\sigma, \sigma^{'}} \cdot \delta \big(\omega-\omega^{'}\big) ~, \nonumber
\end{eqnarray}
\end{widetext}
where $n_{\mathrm{FD}}$ is the Fermi-Dirac distribution function.

In a perturbative RG analysis, we are only interested in the energy range $0 < \omega \ll D$. Consequently in the previous expressions (\ref{SelfEn-2Ord}), (\ref{SelfEn-3Ord}) and (\ref{SelfEn-4Ord}) the terms in powers of $(\omega / D)$ are neglected, and we only keep the factors generating logarithmic contributions in $\ln(\omega / D)$, which are sufficient to derive the two leading order terms of the $\beta$ scaling function as discussed in Section \ref{Obt-Beta}. Whenever necessary, we keep only the principal part of the denominator exhibiting a pole. The integrals in energy are computed in this limit following the method detailed by Gan, \cite{Gan-thesis} we get

\begin{eqnarray}
& & \mbox{Im} \Bigg\{ \int_{-D}^{+D} d \varepsilon_{1} \frac{1}{(\omega -\varepsilon_{1} + \mathrm{i} 0^{+})}  \Bigg\} = - \pi \ ,  
\label{IntEn-2Ord} \\
& & \mbox{Im} \Bigg\{ \int_{-D}^{+D} d \varepsilon_{1} d \varepsilon_{2} \
\frac{n_{\mathrm{FD}}(\varepsilon_{2})}
{(\omega -\varepsilon_{1} + \mathrm{i} 0^{+})(\varepsilon_{1} - \varepsilon_{2}) } \Bigg\} \nonumber \\
&=& \pi \ln(\omega / D) \ ,
\label{IntEn-3Ord} \\
& & \mbox{Im} \Bigg\{ \int_{-D}^{+D}  \
\frac{d \varepsilon_{1} d \varepsilon_{2} d \varepsilon_{3} }
{(\omega -\varepsilon_{1} + \mathrm{i} 0^{+})
(\omega -\varepsilon_{1} +\varepsilon_{2} - \varepsilon_{3} + \mathrm{i} 0^{+}) }  \nonumber \\
& &\hspace{16mm} \times \ \frac{1}{(\varepsilon_{2} - \varepsilon_{3})} \nonumber \\
& &\hspace{16mm} \times \ \Big[ n_{\mathrm{FD}}(-\varepsilon_{1}) n_{\mathrm{FD}}(\varepsilon_{2}) n_{\mathrm{FD}}(-\varepsilon_{3}) \nonumber \\
& &\hspace{22mm} + \ n_{\mathrm{FD}}(\varepsilon_{1}) n_{\mathrm{FD}}(-\varepsilon_{2}) n_{\mathrm{FD}}(\varepsilon_{3}) \Big] \Bigg\} \nonumber \\
&=& \pi \ln(\omega / D) \ .
\label{IntEn-4Ord}
\end{eqnarray}

By incorporating the integrals (\ref{IntEn-2Ord}), (\ref{IntEn-3Ord}) and (\ref{IntEn-4Ord}) into Eqs.\! (\ref{SelfEn-2Ord}), (\ref{SelfEn-3Ord}) and (\ref{SelfEn-4Ord}) we obtain 

\begin{widetext}
\begin{eqnarray}
& & \mbox{Im} \Big[ \mbox{\large{$\Sigma$}}_{ \{ \Gamma \} }^{\sigma, \sigma^{'}}  \big(\omega, \omega^{'}\big) \Big] \nonumber \\
=& & \pi \Bigg( \frac{(N+2S-1)!}{(2S)! (q-1)! (N-q)!} \Big[ 2S \big(N - 1 \big) - q \big(N + 1 \big) +  \big(N^{2}+1 \big) \Big]
\cdot \frac{2S}{4 N} \Bigg) \nonumber \\
&\times& \left(\frac{d_{ \{ \Gamma \} }-1}{\big[d_{ \{ \Gamma \} }\big]^{2} }\right)
\big( J_{\mathrm{K}} \big)^{2} \rho \cdot n_{i}  \cdot \delta_{\sigma, \sigma^{'}} \cdot \delta \big(\omega-\omega^{'}\big)  \cdot
\Bigg\{ \frac{1}{N} - \big( J_{\mathrm{K}} \rho \big) \ln(\omega / D) 
+ \frac{1}{2} \big( J_{\mathrm{K}} \rho \big)^{2} K \cdot \ln(\omega / D) \Bigg\} \ .
\label{SelfEn-final}
\end{eqnarray}
\end{widetext}


\section{Perturbative Renormalization Group approach}
\label{Sec-PRenGr}

In this Section we use a perturbative RG method, the Anderson's ``poor man's scaling'' approach. \cite{Hewson-book,Anderson-PMS} The intermediate coupling (IC) fixed point associated with the model is identified, for various limits characterized by the shape of the   $\{ \Gamma \}$  representation. We start by deriving the $\beta$ scaling function, as a function of $\big(J_{\mathrm{K}}
\rho\big)$. A fixed point corresponds to a zero of the $\beta$ function. A simple analysis of the scaling function properties around a fixed point allows also to determine its stability. \cite{LeBellac} In the case of a stable IC fixed point, a rich physics characterized by a non-Fermi liquid behavior of the system (namely, anomalous temperature dependence of the specific heat, magnetic susceptibility, ...) can be expected.


\subsection{Derivation of the scaling function from the conduction electron self-energy}
\label{Obt-Beta}

In the previous Section \ref{Sub-CalcSelfEn}, we have calculated the first order terms of the conduction electron self-energy in a
perturbative approach. This yields the leading terms of the $\beta$ scaling function
\begin{equation}
\beta ( J_{\mathrm{K}} \rho ) = \frac{d (J_{\mathrm{K}} \rho )}{d (\mathrm{ln}D)}~.
\label{Beta-def}
\end{equation}
The connexion between both results from RG arguments, more precisely the Callan-Symanzik equation \cite{Callan,Symanzik} as explained below.

Without mentioning the spin dependence, we rename the imaginary part of the conduction electron self-energy $\mbox{Im} \Big[ \mbox{\large{$\Sigma$}}_{ \{ \Gamma \} }^{\sigma, \sigma^{'}} \Big]$, Eq.\! (\ref{SelfEn-final}), as  $\mbox{\large{$\Sigma$}}_{ \{ \Gamma \} }^{''}$, to have shorter notation in what follows. Following Gan \textit{et al.}, \cite{Gan-PRL,Gan-thesis} $\mbox{\large{$\Sigma$}}_{ \{ \Gamma \} }^{''}$ can be expressed in the most general way up to the fourth order in perturbation theory as
\begin{eqnarray}
\label{SelfEnRen-4ord}
& &\mbox{\large{$\Sigma$}}_{ \{ \Gamma \} }^{''} \big(\omega, D, J_{\mathrm{K}}\rho \big) \\
&=& 
\frac{a}{\rho} \bigg\{ P_{0}\big( J_{\mathrm{K}} \rho \big)^{2} + P_{1}\ln(\omega / D) \big( J_{\mathrm{K}} \rho \big)^{3}
\nonumber \\
& & \hspace{5mm}
+ \ P_{2}\cdot K \cdot \ln(\omega / D) \big( J_{\mathrm{K}} \rho \big)^{4}  + P_{3}\cdot K \cdot \big( J_{\mathrm{K}} \rho \big)^{4}
\nonumber \\
& & \hspace{5mm} + \ P_{4}\ln^{2}(\omega / D) \big( J_{\mathrm{K}} \rho \big)^{4} + \mathcal{O}\Big[\big(J_{\mathrm{K}} \rho \big)^{5}\Big] \bigg\}~, \nonumber
\end{eqnarray}
where $a$, $P_{0}$, ... , $P_{4}$ are coefficients independent of $\omega$, $D$, $J_{\mathrm{K}}\rho$. $\rho \mbox{\large{$\Sigma$}}_{ \{ \Gamma \} }^{''}$ is the dimensionless scattering rate (i.e., inverse of the quasiparticle lifetime). The key point of the RG theory is that the underlying physics must be independent of the renormalization scheme, namely the cut-off $D$. Therefore the scattering rate is invariant under RG transformation. The Callan-Symanzik equation resulting from this property is then
\begin{eqnarray}
& &\Bigg[\frac{\partial}{\partial (\ln D)} + \beta\big( J_{\mathrm{K}} \rho \big) \frac{\partial}{\partial (J_{\mathrm{K}}\rho)} \Bigg] 
\rho \mbox{\large{$\Sigma$}}_{ \{ \Gamma \} }^{''} \big(\omega, D, J_{\mathrm{K}}\rho \big) \nonumber \\
&=& 0~.
\label{CallSym-eq}
\end{eqnarray}
The $\beta$ scaling function is taken in its most general expansion up to the third order in powers of $J_{\mathrm{K}} \rho$
\begin{eqnarray}
\beta\big( J_{\mathrm{K}} \rho \big) &=& \kappa_{1} \big( J_{\mathrm{K}} \rho \big)^{2} + \kappa_{2} \cdot K \cdot \big( J_{\mathrm{K}} \rho \big)^{3} 
\nonumber \\
& &+ \ \kappa_{3} \big( J_{\mathrm{K}} \rho \big)^{3} + \mathcal{O}\Big[\big(J_{\mathrm{K}} \rho \big)^{4}\Big]~.
\label{BetaF-def}
\end{eqnarray}
 Eqs.\! (\ref{SelfEnRen-4ord}) and (\ref{BetaF-def}) incorporated into Eq.\! (\ref{CallSym-eq}) give
\begin{eqnarray}
& &\big[2 \kappa_{1} P_{0} - P_{1} \big] \big( J_{\mathrm{K}} \rho \big)^{3} + \big[2 \kappa_{2} P_{0} - P_{2} \big] K \cdot \big( J_{\mathrm{K}} \rho \big)^{4} 
\nonumber \\
& &+  \ 2 \kappa_{3} P_{0} \big( J_{\mathrm{K}} \rho \big)^{4} 
+ \big[3 \kappa_{1} P_{1} - 2 P_{4} \big] \ln(\omega / D)  \cdot \big( J_{\mathrm{K}} \rho \big)^{4} \nonumber \\
& &+ \ \mathcal{O}\Big[\big(J_{\mathrm{K}} \rho \big)^{5}\Big] = 0~.
\label{CallSym-dev}
\end{eqnarray}
Eq.\! (\ref{CallSym-dev}) generates logarithmic series valid for any coupling constant $J_{\mathrm{K}}$. Therefore the coefficients in front of the terms of successive orders in $J_{\mathrm{K}}\rho$ are identically 0. The result is
\begin{eqnarray}
\label{BetaF-coeff1}
\kappa_{1} &=& \frac{P_{1}}{2 P_{0}}~, \\
\label{BetaF-coeff2}
\kappa_{2} &=& \frac{P_{2}}{2 P_{0}}~, \\
\label{BetaF-coeff3}
\kappa_{3} &=& 0~, \\
\label{BetaF-coeffP4}
 P_{4} &=& \frac{3}{2} \kappa_{1} P_{1}~.
\end{eqnarray}
The coefficients $P_{0}$, $P_{1}$ and $P_{2}$ are obtained from the imaginary part of the conduction electron self-energy, Eq.\! (\ref{SelfEn-final})
\begin{eqnarray}
\label{SelfEn-coeff0}
P_{0} &=& \frac{1}{N} ~, \\
\label{SelfEn-coeff1}
P_{1} &=& - 1 ~, \\
\label{SelfEn-coeff2}
P_{2} &=& \frac{1}{2} ~.
\end{eqnarray}
We point out here that the coefficients given by Eqs.\! (\ref{SelfEn-coeff0}), (\ref{SelfEn-coeff1}) and (\ref{SelfEn-coeff2}) are exact at this stage for any value $N$, $2S$ and $q$, without any restriction. As far as the consistency equation (\ref{BetaF-coeffP4}) is concerned, it has been checked by Gan  \cite{Gan-thesis} for an impurity spin in the fundamental representation of the SU(2) group. At each order of perturbation our study leads to the same integrals in energy displayed in Eqs.\! (\ref{IntEn-2Ord}), (\ref{IntEn-3Ord}) and (\ref{IntEn-4Ord}). The only difference between Gan's work and our generalized approach is contained in spin factors which depend on the considered group and representation describing the impurity. Since we have calculated these spin factors in an exact way, we conclude that the previous consistency equation is automatically satisfied for any group and any impurity representation we are interested in: the explicit calculation of $P_{4}$ for the generalized $\{ \Gamma \}$ representation is not required.


\subsection{Analysis of the intermediate coupling fixed point}
\label{IC-Analys}

Following the results of previous Section \ref{Obt-Beta}, the  $\beta$ scaling function (\ref{Beta-def}) obtained in a perturbative theory calculation in $J_{\mathrm{K}}$ up to the third order is given by
\begin{eqnarray}
\beta\big( J_{\mathrm{K}} \rho \big) &=& -\frac{N}{2} \big( J_{\mathrm{K}} \rho \big)^{2} 
+ \frac{NK}{4} \big( J_{\mathrm{K}} \rho \big)^{3} \nonumber \\
& &+ \ \mathcal{O}\Big[\big(J_{\mathrm{K}} \rho \big)^{4}\Big].
\label{Beta-result}
\end{eqnarray}
One of the main results of this work is that Eq.\! (\ref{Beta-result}) is valid for any value of the parameters $2S$ and $q$ characterizing the L-shaped representation of the impurity spin, provided that the perturbative expansion in $J_{\mathrm{K}}$ is valid as discussed in the next paragraph.

The critical value $J^{*}$ taken by the Kondo coupling at the IC fixed point is obtained from the zero of the $\beta$ function, i.e., $\beta\big( J^{*} \rho \big) = 0$. We get
\begin{eqnarray}
J^{*} \rho = \frac{2}{K} + \mathcal{O}\left[\frac{1}{K^{2}} \right]~.
\label{J-IntFix}
\end{eqnarray}
The latter expression for $J^{*}$ allows one to determine the domain of validity of the perturbative expansion. The RG perturbative approach we use is valid in the weak coupling regime only. The IC fixed point has to be in the vicinity of the free electron fixed point of the system. Consequently, the perturbative expansion of the $\beta$ function is trustworthy exclusively when $J^{*}$ is close to 0, which means by using Eq.\! (\ref{J-IntFix}) $K \gg 1$. Since each conduction electron loop which gives a factor of ($-K$) contains two interacting vertices at least, and each term $J^{*} \rho$ contributes by a factor of $1/K$, we conclude that the diagrams displayed in Figs.\! \ref{SelfEnDiags-Fig2}, \ref{SelfEnDiags-Fig3}, and \ref{SelfEnDiags-Fig4} include all the contributions to the electronic self-energy up to the order $\mathcal{O}\big[1/K^{3}\big]$ at the IC fixed point. This argument allows one to check the self-consistency of the calculation. Another restriction is due to the domain of validity of the perturbative expansion. As previously noticed by Abrikosov-Migdal \cite{AbriMig} and Affleck-Ludwig, \cite{AffLud-NPB1991} the fourth order term in the expansion of the $\beta$ function, $\mathcal{O}\Big[\big( J_{\mathrm{K}} \rho \big)^{4}\Big]$, contains a coefficient proportional to the eigenvalue of the quadratic Casimir operator (\ref{Casi-Gam-red}) related to the impurity spin representation. In order that the latter contribution which has not been taken into account here remains negligible  compared to the second and third order contributions, we must also suppose that $N$, $2S$ and $q$ are small compared to $K$. To summarize, the domain of validity of the scaling function perturbative expansion is given by 
\begin{equation}
K \gg 1, ~N \ll K, ~2S \ll K, ~q \ll K~.
\label{PRG-limit}
\end{equation}
Consequently the IC fixed point describes an overscreened impurity, as discussed originally by Nozi\`eres and Blandin. \cite{BlandNoz-1980} The single-channel case turns out to be out of reach of the present study.

In addition to the value of $J^{*}$, another key quantity which can be drawn from the $\beta$ function is its slope $\Delta$ at the IC fixed point, namely
\begin{equation}
\Delta = \left. \frac{d \beta}{d (J_{\mathrm{K}} \rho )} \right]_{J_{\mathrm{K}} = J^{*}}
= \frac{N}{K} + \mathcal{O}\left[\frac{1}{K^{2}} \right]~.
\label{DeltaUniv}
\end{equation}
In the limit given by Eq.\! (\ref{PRG-limit}) that we consider, $\Delta$ is automatically positive therefore the IC fixed point is stable. \cite{LeBellac} The slope $\Delta$ determines the critical exponent of the model. It is a strong mark of the universality concept characteristic of the Kondo effect in the low-energy regime. \cite{Loram-1970,Andrei-RMP1983} This regime describes the system at temperatures that are low compared to the bandwidth $D$. In that case the physics of the system is governed by a stable IC fixed point, and the temperature dependence of the various physical quantities at low $T$ obeys a power law in $T / T_{\mathrm{K}}$ which is entirely controlled by $\Delta$ ($T_{\mathrm{K}}$ being the Kondo temperature).

The Kondo temperature is the only relevant energy scale in the problem, it is generated dynamically by the interactions with the impurity spin. It plays an essential role in the Kondo effect: $T_{\mathrm{K}}$  characterizes a crossover between the weak coupling regime and the low-energy regime. This temperature scale is invariant under the scaling process. One can easily derive it by inverting and integrating the $\beta$ function expression. \cite{Hewson-book} By using Eqs.\! (\ref{Beta-def}) and (\ref{Beta-result}) we get
\begin{equation}
k_{\mathrm{B}} T_{\mathrm{K}} \sim D (J_{\mathrm{K}} \rho)^{K/N} \exp{\bigg[{-\frac{2}{N(J_{\mathrm{K}} \rho)}}\bigg]}~.
\label{KondoTemp}
\end{equation}

\begin{figure}[t]
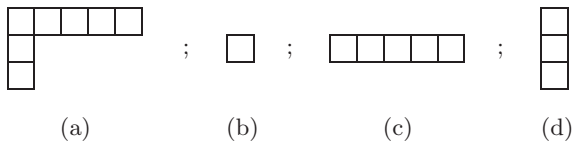

\begin{eqnarray*} 
& &\yng(5,1,1)~~~~~;~~~~\yng(1)~~~~;~~~~\yng(5)~~~~;~~~~\yng(1,1,1) \\ \\
& &\hspace{7mm} \mathrm {(a)} \hspace{18mm} \mathrm {(b)} \hspace{16.5mm} \mathrm {(c)} \hspace{17mm} \mathrm {(d)} 
\end{eqnarray*}
\caption {Irreducible SU($N$) representations of the impurity spin considered in the present work:
(a) L-shaped ; (b) fundamental, $2S=q=1$ ; (c) symmetric, $q=1$ ; (d) antisymmetric, $2S=1$.} 
\label{ImpSpinRep-Fig}
\end{figure}
As mentioned above, the results we have obtained are valid in the overscreened regime for any value of the parameters  $N$, $2S$ and $q$, provided that Eq.\! (\ref{PRG-limit}) is satisfied. The impurity spin representations that are adressed by the present study are displayed in Fig.\! \ref{ImpSpinRep-Fig}. We conclude that {all the Kondo models defined by these different impurity spin representations belong to the same universality class in the overscreened regime, which is characterized by the following low-temperature behaviors \cite{Gan-thesis}
\begin{eqnarray}
&-&\mathrm{resistivity~due~to~Kondo~scattering:} \nonumber \\
& &~~\delta \rho_{\mathrm{e}}(T) = \rho_{\mathrm{e}}(T=0) - \alpha
\bigg( \frac{T}{T_{\mathrm{K}}} \bigg)^{\Delta} ~, \nonumber \\
&-& \mathrm{impurity~specific~heat~:}~C_{\mathrm{imp}}(T) \sim \bigg( \frac{T}{T_{\mathrm{K}}} \bigg)^{2 \Delta}
 ~, \nonumber \\
&-& \mathrm{magnetic~susceptibility~:}~\chi_{\mathrm{imp}}(T) \sim \frac{1}{T}
\bigg( \frac{T}{T_{\mathrm{K}}} \bigg)^{2 \Delta} ~. \nonumber
\end{eqnarray}
We can notice that the system described by an IC fixed point exhibits non-Fermi liquid properties. As in the original single-channel Kondo model characterized by an exactly screened spin 1/2 impurity, the resistivity increases when the temperature is lowered. Denoting by $\vec{\bm{S}}_{ef\!f}$ the effective spin of the impurity screened by the conduction electrons and $\langle ~ \rangle$ the thermal average, we have \cite{Hewson-book} 
$\left\langle \big( \vec{\bm{S}}_{ef\!f} \big)^{2} \right\rangle \sim T \chi_{\mathrm{imp}}(T) \sim T^{2 \Delta} \rightarrow 0$ 
as $T \rightarrow 0$ which indicates that the IC fixed point is associated with a ground state corresponding to a spin singlet. In the overscreened regime the impurity spin is completely quenched by conduction electrons.

The value of the universal critical exponent $\Delta$ we have obtained in Eq.\! (\ref{DeltaUniv}) is in agreement with the result of Gan \textit{et al.} who considered a spin 1/2 in the SU(2) group. \cite{Gan-PRL,Gan-thesis} The case of larger spins ($S > 1/2$) in the SU(2) group has also been studied, \cite{FabriZar-1996,Shaw-1998} leading to similar results. Eq.\! (\ref{DeltaUniv}) allows one to recover the overscreened limit  of the exact solution of the multichannel Kondo model, obtained by using the Bethe ansatz for the symmetric impurity spin representations of SU($N$), \cite{jaz} and boundary conformal field theory in the case of an arbitrary impurity spin representation of the SU(2) \cite{AffLud-NPB1991} and  SU($N$) \cite{AffLud-1994} groups. It also corroborates the large-$N$ studies of the SU($N$)$\times$SU($K$) Kondo model involving either the fundamental \cite{CoRu} or the antisymmetric \cite{pgks} impurity spin representations, although the $K \sim N$ limit considered in these works is different from ours.


\section{Summary and conclusions}
\label{Conclu}

In this paper, we have studied a multichannel SU($N$) Kondo model, involving a mixed symmetry impurity spin representation. Namely the  generalized impurity spin is realized by a combination of $2S$ bosonic and $q$ fermionic degrees of freedom, in the presence of $K$ conduction electron channels. The study has been carried out by making use of a perturbative renormalization group technique, the Anderson ''poor man's scaling'' approach. The main difficulty for the description of spin systems is that spin operators obey neither Fermi nor Bose statistics, leading to the absence of Wick's decomposition theorem in this context. Consequently conventional perturbative methods are prohibited. We have used the Abrikosov pseudofermion formulation to describe the impurity spin states, and tackled the problem by making use of a method initiated by Popov and Fedotov. Its basic principle consists in introducing a set of auxiliary chemical potentials which eliminates the unphysical Hilbert sub-space keeping the physical spin states only, and therefore enables a proper handling of the fermionic constraint.

The partition function has been obtained within a path integral approach. The general expression of the conduction electron self-energy imaginary part has been derived perturbatively in powers of the Kondo coupling constant, i.e., by considering the weak coupling limit. At each order of a perturbative calculation the derivation involves two steps, namely the calculation of the spin factor and the computation of the fermionic propagator contribution which includes the Popov-Fedotov auxiliary chemical potentials. The imaginary part of the conduction electron self-energy is proportional to the electronic scattering rate, which is a physical quantity and therefore is invariant under renormalization group transformation. The Callan-Symanzik equation can be applied, and yields the derivation of the two leading terms of the $\beta$ scaling function, i.e., up to the order $\mathcal{O}\big[1/K^{2}\big]$. We have established that this truncated expansion of the $\beta$ function is valid in the overscreened (Nozi\`eres - Blandin) regime, for any value of  $N$, $2S$ and $q$ providing that these parameters remain small compared with $K$. The intermediate coupling (IC) fixed point describing the low-energy physics of the model has been identified in the overscreened regime. The critical exponent $\Delta$, corresponding to the slope of the $\beta$ function towards $J_{\mathrm{K}} \rho$ at the IC fixed point, is given by
\begin{equation}
\Delta = \frac{N}{K} + \mathcal{O}\left[\frac{1}{K^{2}} \right]~. \nonumber
\end{equation}
$\Delta$ governs the temperature dependence of the various physical quantities at low $T$ which are of non-Fermi liquid type, as discussed in Section \ref{Sec-PRenGr}. The IC fixed point describes the system in its ground state which is a spin singlet: in the overscreened regime the impurity spin is completely quenched by the conduction electrons located on its site.

Several types of impurity spin representations have been considered in previous works and can be viewed as special limits of the mixed symmetry formulation we have used. Hence direct comparisons with our study have been made possible. Our results are in agreement with those obtained by other methods, like Bethe ansatz and boundary conformal field theory. We have established that the Kondo models defined successively by fundamental, symmetric, antisymmetric and mixed symmetry impurity spin representations yield exactly the same low-energy physics in the overscreened regime. To summarize the present work gives a unified view of the overscreened Kondo regime, for a large set of representations of the impurity spin. 
It is also possible to consider arbitrary impurity spin representations, beyond the cases mentioned here. For example one can think about representations of the SU($N$) group with rectangular-shaped Young tableaux, as discussed in Ref.\! \onlinecite{Natan-1998}. The method we present can be generalized to study the case of an arbitrary SU($N$) representation. No additional conceptual difficulty has to be expected. The spin factors associated with such a representation, which are present in the conduction electron self-energy, have to be computed following the derivation displayed in Appendix \ref{Appen-SpinFact}. The traces over $\tilde{\bm{\Gamma}}$ matrices we have to consider involve products of two or three generators in the impurity spin representation, and therefore the spin factors can be derived exactly for any irreducible representation of SU($N$). \cite{Okubo-1982,Schell-1999,Okubo-1977-2} However the computation of these spin factors as a function of the Young tableau parameters describing a completely arbitrary impurity spin representation certainly requires extremely lengthy algebra, as the complexity of Eqs.\!  (\ref{TrGam3-fin}), (\ref{Contrib-Spin3}) and (\ref{Contrib-Spin3b}), which are restricted to the L-shaped case we consider, suggests it easily. \\
\indent Finally we mention the case of the single-channel SU($N$) Kondo model involving a mixed symmetry impurity spin, which we considered in a previous work. \cite{JLB-Kondo03,JLB-Kondo03Short} In such a case the L-shaped impurity exhibits a rather surprising duality: it is exactly screened if one considers the fermionic component, but underscreened towards the bosonic one. An interesting problem remains. Namely, the understanding of the low-energy physics in the regime characterized by an instability of the strong coupling fixed point turns out to be out of reach of the present work. This issue deserves further investigations. It might have important future applications for the Kondo lattice problem, \cite{Doniach-1977} with potential implications for the description of heavy-fermion compounds. In order to gain some insight it would be interesting to carry out a non-perturbative study. Unfortunately it is quite a difficult task. Indeed well-known methods usually efficient to study the Kondo problem, as numerical renormalization group or Bethe ansatz techniques, have severe limitations in the context of the mixed symmetry representation. For example, the simplest impurity that displays an IC fixed point in the single-channel case implies a large Hilbert space which restricts the efficiency of a numerical renormalization group study. The difficulties encountered by the Bethe ansatz are of different nature: due to the mixed symmetry properties of the impurity spin representation, the model with Kondo coupling only is not integrable. \cite{Natan-PCom} 


\begin{acknowledgements}

We would like to thank Natan Andrei for highly valuable comments and discussions at several steps of this work. We are grateful to Marijana Kir\'{c}an for a careful reading of the manuscript. D.B. would like to acknowledge extremely fruitful discussions with Alex Hewson, Naoto Nagaosa, Philippe Nozi\`eres and Roland Zeyher.

\end{acknowledgements}


\appendix


\section{Properties of the SU($N$) generators}
\label{Appen-SUNgener}

In this Appendix we give some properties of the generators $\big\{ \tilde{\bm{\sigma}} \big\}$ and $\big\{ \tilde{\bm{\Gamma}} \big\}$, in the case of the two types of irreducible representations we have to consider in the SU($N$) group. The first one concerns the conduction electron, it is the fundamental (single-box) representation $\{ \sigma \}$ displayed in Fig.\! \ref{FundaRep-Fig}. Its dimension is
\begin{eqnarray}
d_{ \{ \sigma \} } = N,
\label{Dim-sig}
\end{eqnarray}
and the eigenvalue of the quadratic Casimir operator corresponding to $\{ \sigma \}$ is
\begin{eqnarray}
\mathcal{C}_{2}\big(\{ \sigma \} \big) = \frac{N^{2}-1}{2N},
\label{Casi-sig}
\end{eqnarray}
which in the case of a spin $S$ in SU($2$) takes the usual value $S(S+1)$. The associated traceless hermitian matrices $\big\{ \tilde{\bm{\sigma}}^{A} \big\}_{1\leq A \leq N^{2}-1}$ obey the following properties \cite{GreiMull}
\begin{eqnarray}
\label{Mult-law}
&-& \mbox{multiplication law: } \nonumber \\
& &\tilde{\bm{\sigma}}^{A} \tilde{\bm{\sigma}}^{B} = 
\frac{1}{2N} \delta_{AB} \tilde{\bm{I}}_{\{ \sigma \} }  \nonumber \\
& &\hspace{14mm}+ \ \frac{1}{2} \sum_{C=1}^{N^{2}-1} \left(d^{ABC} + \mathrm{i} f^{ABC}\right) \tilde{\bm{\sigma}}^{C}~, ~~ \\
\label{Norm-sig}
&-& \mbox{normalization relation: } \nonumber \\
& &\mathrm{Tr}\big[ \tilde{\bm{\sigma}}^{A} \tilde{\bm{\sigma}}^{B} \big] = \frac{1}{2} \delta_{AB}~, \\
\label{Compl-sig}
&-& \mbox{completeness relation: } \nonumber \\
& &\sum_{A=1}^{N^{2}-1} \bm{\sigma}_{\alpha,\beta}^{A}  \bm{\sigma}_{\gamma,\delta}^{A} =
\frac{1}{2} \left( \delta_{\alpha,\delta} \delta_{\beta,\gamma} - \frac{1}{N} \delta_{\alpha,\beta} \delta_{\gamma,\delta} \right)~, ~~~~~
\end{eqnarray}
where $A$, $B$ and $C$ label the SU($N$) generators matrices ($1 \leq A, B, C \leq N^{2}-1$),  $\tilde{\bm{I}}_{\{ \sigma \} }$ is the ($N \times N$) identity matrix, $\alpha$, $\beta$, $\gamma$ and $\delta$ label electronic spin colors ($1 \leq \alpha, \beta, \gamma, \delta \leq N$), and $d^{ABC}$ and $f^{ABC}$ are the structure constants related to the $\tilde{d}$ and $\tilde{f}$ Gell-Mann tensors, respectively. \cite{GellMann-1962} The structure constants are characteristic of the underlying Lie algebra. These tensors follow the rules
\begin{eqnarray}
&-& \mbox{$\tilde{d}$ is totally symmetric: \ \ } d^{ABC} = d^{BAC}~, \nonumber \\ 
&-& \mbox{$\tilde{f}$ is totally antisymmetric: \ \ } f^{ABC} = -f^{BAC}~, \nonumber \\
&-& \mbox{$\tilde{d}$ and $\tilde{f}$ are traceless tensors: }  \nonumber \\
& &\sum_{B} d^{ABB} = \sum_{B} f^{ABB} = 0~. \nonumber
\end{eqnarray}
In SU($2$), $\tilde{d}$ = 0 and $\tilde{f}$ = $\tilde{\epsilon}$, the totally antisymmetric (Levi-Civitta) tensor of rank 3.
In further calculations, we use the following identities \cite{dfTENS-1967,dfTENS-1998}
\begin{eqnarray}
\label{Multdd}
& &\sum_{B,C} d^{ACB} d^{GBC} = \frac{N^{2}-4}{N} \delta_{AG} ~, \\
\label{Multff}
& &\sum_{B,C} f^{ACB} f^{GBC} = - N \delta_{AG} ~, \\
\label{Multdf}
& &\sum_{B,C} d^{ACB} f^{GBC} = 0 ~.
\end{eqnarray}

The second representation of interest deals with the impurity spin. It is a L-shaped representation denoted by $\{ \Gamma \}$, and characterized by $2S$ boxes in the first row (horizontal line) and $q$ boxes in the first column (vertical line), as displayed in Fig.\! \ref{LShape-Fig}. Its dimension can be derived by the use of Robinson's formula \cite{IVerona}
\begin{eqnarray}
d_{ \{ \Gamma \} } = \frac{2S}{2S+q-1} C_{N+2S-1}^{2S} C_{N-1}^{q-1}~,
\label{Dim-Gam}
\end{eqnarray}
and the eigenvalue of the quadratic Casimir operator associated with $\{ \Gamma \}$ is given by \cite{Okubo-1977}
\begin{eqnarray}
\mathcal{C}_{2}\big(\{ \Gamma \} \big) &=& 
\frac{1}{2}\Bigg\{ \frac{(2S+q-1)\big[N^{2}-(2S+q-1)\big]}{N}  \nonumber \\
& &  \hspace{4mm} + \ \sum_{j=1}^{N}m_{j}(m_{j}+1-2j)\Bigg\}~, 
\label{Casi-Gam-full}
\end{eqnarray}
where $m_{j}$ is the number of boxes in the $j$-th row ($1 \leq j \leq N$) of the Young tableau describing $\{ \Gamma \}$. Finally
\begin{eqnarray}
& &\mathcal{C}_{2}\big(\{ \Gamma \} \big)  \nonumber \\
&=& \frac{2S+q-1}{2}\left( 2S-q+N - \frac{2S+q-1}{N}\right)~.~~~~~~~~~
\label{Casi-Gam-red}
\end{eqnarray}
We give now useful identities satisfied by the $\big( d_{ \{ \Gamma \} } \times d_{ \{ \Gamma \} }\big)$ components of the $\big\{ \tilde{\bm{\Gamma}}^{A} \big\}$ matrices. The normalization relation is \cite{Okubo-1982}
\begin{eqnarray}
\mathrm{Tr}\big[ \tilde{\bm{\Gamma}}^{A} \tilde{\bm{\Gamma}}^{B} \big] 
&=& \frac{d_{ \{ \Gamma \} } \mathcal{C}_{2}\big(\{ \Gamma \} \big)}{d_{ \{ \sigma \} } \mathcal{C}_{2}\big(\{ \sigma \} \big)} 
\mathrm{Tr}\big[ \tilde{\bm{\sigma}}^{A} \tilde{\bm{\sigma}}^{B} \big]  \nonumber \\
&=& \frac{d_{ \{ \Gamma \} } \mathcal{C}_{2}\big(\{ \Gamma \} \big)}{d_{ \{ \sigma \} } \mathcal{C}_{2}\big(\{ \sigma \} \big)}
\frac{\delta_{AB}}{2}~.
\label{Norm-Gam}
\end{eqnarray}
The generators in the $\{ \Gamma \}$ representation obey also some particular multiplication laws for the contracted indices close to each other 
\cite{Schell-1999}
\begin{eqnarray}
\label{ProdGam-3}
& &\sum_{B} \tilde{\bm{\Gamma}}^{B} \tilde{\bm{\Gamma}}^{A} \tilde{\bm{\Gamma}}^{B} = 
\left\{ \mathcal{C}_{2}\big(\{ \Gamma \} \big) - \frac{\mathcal{C}_{2}\big(\{ \mbox{Adj} \} \big)}{2} \right\} \tilde{\bm{\Gamma}}^{A}~,~~~~~~~~ \\
\label{ProdGam-4}
& &\sum_{B}   \tilde{\bm{\Gamma}}^{C} \tilde{\bm{\Gamma}}^{B} \tilde{\bm{\Gamma}}^{B} \tilde{\bm{\Gamma}}^{A} =
\mathcal{C}_{2}\big(\{ \Gamma \} \big) \tilde{\bm{\Gamma}}^{C} \tilde{\bm{\Gamma}}^{A}~,~~~~
\end{eqnarray}
where $\mathcal{C}_{2}\big(\{ \mbox{Adj} \} \big)$ is the eigenvalue of the quadratic Casimir operator associated with the adjoint representation of SU($N$). Due to the choice of normalization given by Eq.\! (\ref{Norm-sig}) we have: 
$\mathcal{C}_{2}\big(\{  \mbox{Adj} \} \big) = N$.


\section{Popov and Fedotov method}
\label{Appen-Popov}

In this Appendix we briefly explain a method pioneered by Popov and Fedotov, \cite{PopovFed} in order to handle the constraint (\ref{Cons-fermi}) fixing the number of pseudofermions on the impurity site. Originally formulated for spins 1/2 and 1 of the SU(2) group, this powerful method has been developed for arbitrary spin in SU(2) \cite{Opper1994} and generalized to SU($N$). \cite{Kiselev2001} It has already been used to study the Heisenberg model, \cite{TejiOgu1995} the Kondo lattice \cite{BouiKis1999} and disordered systems. \cite{KisOpp-JETP2000,KisOpp-PRL2000}

The principle is to express the partition function of a spin system in terms of partition functions of a corresponding fermionic system, pseudofermions being used to describe the spin states. The problem is to incorporate the fermionic constraint properly, namely to eliminate the unphysical Hilbert sub-space without removing the true spin states. This can be done by noticing that the Kondo hamiltonian (\ref{HK-full}) without imposing any constraint is disconnected in the peudofermion charge sectors. Consequently the trace over the full Fock space (i.e., without constraint) can be rewritten by applying the Pauli principle as
\begin{eqnarray}
& &\mathrm{Tr}_{\mbox{\scriptsize{Fock}}} \mathrm{exp} \Big[- \beta \big( \mathcal{H}_{\mathrm{K}} - \mu \hat{n}_{f} \big) \Big]  \nonumber \\
&=& \sum_{n_{f}=0}^{d_{ \{ \Gamma \} } } \mathrm{Tr}_{n_{f}} \mathrm{exp} \Big[- \beta \big( \mathcal{H}_{\mathrm{K}} - \mu \hat{n}_{f} \big) \Big]~,
 \label{TrFock-detail}
\end{eqnarray}
where $\mathrm{Tr}_{n_{f}}$ is the local subtrace over the states with a fixed number $n_{f}$ of pseudofermion(s) on the impurity site. The physical partition function $\mathcal{Z}$ (\ref{Tr-PartFu}), which includes the fermionic constraint, is together with Eq.\! (\ref{TrFock-detail})
\begin{eqnarray}
\mathcal{Z} = \delta \big( n_{f} - 1 \big) \cdot \mathrm{Tr}_{\mbox{\scriptsize{Fock}}} \mathrm{exp} \Big[- \beta \big( 
\mathcal{H}_{\mathrm{K}} - \mu \hat{n}_{f} \big) \Big] ~.
\label{PartFu-TrDelta}
\end{eqnarray}
By using the identity
\begin{eqnarray}
\delta \big( n_{f} -1 \big)= \frac{1}{d_{ \{ \Gamma \} }} \sum_{n=1}^{d_{ \{ \Gamma \} } }
 \mathrm{e}^{\ \mathrm{i} \pi (2n-1) / d_{ \{ \Gamma \} }} \mathrm{e}^{\ \mathrm{i} \pi (1-2n) n_{f} / d_{ \{ \Gamma \} }}, \nonumber 
\label{Delta-ident}
\end{eqnarray}
incorporated into Eq.\! (\ref{PartFu-TrDelta}) we obtain
\begin{eqnarray}
\mathcal{Z} = \int \mathcal{D}\mu \ P\{ \mu \} \ \mathrm{Tr}_{\mbox{\scriptsize{Fock}}} \mathrm{exp} 
\Big[- \beta \big( \mathcal{H}_{\mathrm{K}} - \mu \hat{n}_{f} \big) \Big]~,
\label{PartFu-TrSumDelta}
\end{eqnarray}
with
\begin{eqnarray}
P\{ \mu \} &=& \frac{1}{d_{ \{ \Gamma \} }} \sum_{n=1}^{d_{ \{ \Gamma \} } }  
\mathrm{exp} \left[\frac{\mathrm{i} \pi}{d_{ \{ \Gamma \} }} (2n-1) \right] \nonumber \\
& & \hspace{14mm} \times \ \delta \left( \mu +  \frac{\mathrm{i} \pi}{\beta d_{ \{ \Gamma \} }} (2n-1) \right)~,~~~
\label{Def-PmuAppen}
\end{eqnarray}
which is exactly Eq.\! (\ref{Def-Pmu}).


\section{Derivation of the free propagators}
\label{Appen-FreeProp}

In order to compute the conduction electron self-energy at the first orders in a perturbative theory in $J_{\mathrm{K}}$, the first step consists in deriving the free propagators associated with the conduction electrons and the pseudofermions. It is performed in this Appendix. Due to the presence of the imaginary chemical potentials (\ref{Def-Pmu}) in the partition function $\mathcal{Z}$ (\ref{LnHK-PartFu}), we should determine the free propagators carefully. For that purpose we use the technique of the generating function. \cite{NegOr,Tsvelik-book1}

Let us denote by $\mathrm{J}^{*}$, $\mathrm{J}$ and $\eta^{*}$, $\eta$ the external sources coupled to the conduction electron $\bar{c}$, $\bar{c}^{*}$ and impurity fermion $\bar{f}$, $\bar{f}^{*}$ variables, respectively. These external sources are Grassmann numbers. The generating function is
\begin{widetext}
\begin{eqnarray}
\label{GenerFunc}
& &\mathcal{Z}^{G} [\mathrm{J}, \mathrm{J}^{*}, \eta, \eta^{*}] \\
&=& \int \mathcal{D}\bar{c}^{*} \mathcal{D}\bar{c} \mathcal{D}\bar{f}^{*} \mathcal{D}\bar{f} 
\Bigg(
\frac{1}{d_{ \{ \Gamma \} }} \  \sum_{n=1}^{d_{ \{ \Gamma \} } }  \mathrm{exp} \left[\mathrm{i} \beta \mu_{n} \right] 
\cdot \mathrm{exp} \Big[
-  \sum_{\mathrm{i} \omega_{m}} \sum_{\bm{k},\sigma,\lambda} 
\big\{ \bar{c}_{\bm{k},\sigma,\lambda}^{*}(\mathrm{i} \omega_{m})(-\mathrm{i} \omega_{m} +  \varepsilon _{\bm{k}}) 
\bar{c}_{\bm{k},\sigma,\lambda}(\mathrm{i} \omega_{m}) \nonumber \\
& & \hspace{88mm} - \mathrm{J}_{\bm{k},\sigma,\lambda}^{*}(\mathrm{i} \omega_{m}) \bar{c}_{\bm{k},\sigma,\lambda}(\mathrm{i} \omega_{m})
- \bar{c}_{\bm{k},\sigma,\lambda}^{*}(\mathrm{i} \omega_{m}) \mathrm{J}_{\bm{k},\sigma,\lambda}(\mathrm{i} \omega_{m}) \}  \nonumber \\
& &\hspace{69mm} - \sum_{\mathrm{i} \omega_{m}} \sum_{\gamma} \big\{
\bar{f}_{\gamma}^{*}(\mathrm{i} \omega_{m}) (-\mathrm{i} \omega_{m} + \mathrm{i}\mu_{n})
\bar{f}_{\gamma}(\mathrm{i} \omega_{m}) \nonumber \\
& & \hspace{85mm} - \eta_{\gamma}^{*}(\mathrm{i} \omega_{m}) \bar{f}_{\gamma}(\mathrm{i} \omega_{m})
 - \bar{f}_{\gamma}^{*}(\mathrm{i} \omega_{m}) \eta_{\gamma} (\mathrm{i} \omega_{m}) \big\}
\Big] \Bigg)~. \nonumber
\end{eqnarray}
If each source is put to 0 one can see immediately that $\mathcal{Z}^{G}$ is identical to the free partition function $\mathcal{Z}_{0}$, describing the system in the absence of the Kondo interaction. The free propagators are given by the thermal averages
\begin{eqnarray}
\label{FreePrEl-DJ}
\langle \bar{c}_{\bm{k}_{\bm{0}},\sigma_{0},\lambda_{0}}(\mathrm{i} \omega_{m_{0}})
\bar{c}_{\bm{k}_{\bm{0}},\sigma_{0},\lambda_{0}}^{*}(\mathrm{i} \omega_{m_{0}}) \rangle_{0} &=&
\left.
 \frac{1}{\mathcal{Z}_{0}} 
\frac{\partial^{2}\mathcal{Z}^{G} [\mathrm{J}, \mathrm{J}^{*}, \eta, \eta^{*}]}  
{\partial \mathrm{J}_{\bm{k}_{\bm{0}},\sigma_{0},\lambda_{0}} (\mathrm{i} \omega_{m_{0}})
\partial \mathrm{J}_{\bm{k}_{\bm{0}},\sigma_{0},\lambda_{0}}^{*} (\mathrm{i} \omega_{m_{0}})}
\right]_{\begin{array}{l} \mbox{\scriptsize{$\mathrm{J} = \mathrm{J}^{*} = 0$}} \\ 
 \mbox{\scriptsize{$\eta = \eta^{*} = 0$}} \end{array}} , \ \\ 
\label{FreePrFer-Deta}
\langle \bar{f}_{\gamma_{0}}(\mathrm{i} \omega_{m_{0}}) \bar{f}_{\gamma_{0}}^{*}(\mathrm{i} \omega_{m_{0}})  \rangle_{0}  &=&
\left.
 \frac{1}{\mathcal{Z}_{0}} 
\frac{\partial^{2}\mathcal{Z}^{G} [\mathrm{J}, \mathrm{J}^{*}, \eta, \eta^{*}]}  
{\partial \eta_{\gamma_{0}} (\mathrm{i} \omega_{m_{0}}) \partial \eta_{\gamma_{0}}^{*}(\mathrm{i} \omega_{m_{0}})}
\right]_{\begin{array}{l} \mbox{\scriptsize{$\mathrm{J} = \mathrm{J}^{*} = 0$}} \\ 
 \mbox{\scriptsize{$\eta = \eta^{*} = 0$}} \end{array}}.
\end{eqnarray}
We use then the following identity for Gaussian integrals over Grassmann fields
\begin{eqnarray}
& &\int \mathcal{D}\bar{c}^{*} \mathcal{D}\bar{c} \mathcal{D}\bar{f}^{*} \mathcal{D}\bar{f} 
\mathrm{exp} \Big[
-  \sum_{\mathrm{i} \omega_{m}} \sum_{\bm{k},\sigma,\lambda} 
\big\{ \bar{c}_{\bm{k},\sigma,\lambda}^{*}(\mathrm{i} \omega_{m})(- \mathrm{i} \omega_{m} +  \varepsilon _{\bm{k}}) 
\bar{c}_{\bm{k},\sigma,\lambda}(\mathrm{i} \omega_{m}) \nonumber \\
& & \hspace{52mm} - \mathrm{J}_{\bm{k},\sigma,\lambda}^{*}(\mathrm{i} \omega_{m}) \bar{c}_{\bm{k},\sigma,\lambda}(\mathrm{i} \omega_{m})
- \bar{c}_{\bm{k},\sigma,\lambda}^{*}(\mathrm{i} \omega_{m}) \mathrm{J}_{\bm{k},\sigma,\lambda}(\mathrm{i} \omega_{m}) \}  \nonumber \\
& &\hspace{34mm} - \sum_{\mathrm{i} \omega_{m}} \sum_{\gamma} \big\{
\bar{f}_{\gamma}^{*}(\mathrm{i} \omega_{m}) (-\mathrm{i} \omega_{m} + \mathrm{i}\mu_{n})
\bar{f}_{\gamma}(\mathrm{i} \omega_{m}) \nonumber \\
& & \hspace{50mm} - \eta_{\gamma}^{*}(\mathrm{i} \omega_{m}) \bar{f}_{\gamma}(\mathrm{i} \omega_{m})
 - \bar{f}_{\gamma}^{*}(\mathrm{i} \omega_{m}) \eta_{\gamma} (\mathrm{i} \omega_{m}) \big\} \Big]  \nonumber \\
&=& \ \ \mathrm{exp} \Bigg[ \ \ \ \sum_{\mathrm{i} \omega_{m}} \sum_{\bm{k},\sigma,\lambda} 
\Bigg\{ \mathrm{J}_{\bm{k},\sigma,\lambda}^{*}(\mathrm{i} \omega_{m})
\Bigg(\frac{-1}{\mathrm{i} \omega_{m} - \varepsilon _{\bm{k}}} \Bigg)
\mathrm{J}_{\bm{k},\sigma,\lambda}(\mathrm{i} \omega_{m})  \Bigg\} \nonumber \\
& & \hspace{14mm} + \ \sum_{\mathrm{i} \omega_{m}} \sum_{\gamma} 
\Bigg\{ \eta_{\gamma}^{*}(\mathrm{i} \omega_{m})\Bigg(\frac{-1}{\mathrm{i} \omega_{m} - \mathrm{i}\mu_{n}} \Bigg)
 \eta_{\gamma} (\mathrm{i} \omega_{m}) \Bigg\} 
\Bigg] \nonumber \\
& & \times \int  \mathcal{D}\bar{c}^{*} \mathcal{D}\bar{c} \mathcal{D}\bar{f}^{*} \mathcal{D}\bar{f}
\mathrm{exp} \Bigg[ 
-  \sum_{\mathrm{i} \omega_{m}} \sum_{\bm{k},\sigma,\lambda} 
\big\{ \bar{c}_{\bm{k},\sigma,\lambda}^{*}(\mathrm{i} \omega_{m})(-\mathrm{i} \omega_{m} +  \varepsilon _{\bm{k}}) 
\bar{c}_{\bm{k},\sigma,\lambda}(\mathrm{i} \omega_{m}) \big\} \nonumber \\
& & \hspace{38mm} - \sum_{\mathrm{i} \omega_{m}} \sum_{\gamma} \big\{
\bar{f}_{\gamma}^{*}(\mathrm{i} \omega_{m}) (-\mathrm{i} \omega_{m} + \mathrm{i}\mu_{n})
\bar{f}_{\gamma}(\mathrm{i} \omega_{m}) \big\} \Bigg]~.
\label{GaussId}
\end{eqnarray}
By using Eq.\! (\ref{GaussId}) one can easily obtain the derivatives of the generating function $\mathcal{Z}^{G}$ (\ref{GenerFunc}) with respect to the sources, allowing us to get the thermal averages
\begin{eqnarray}
\label{FreePrEl-DJ2}
& & \left.
\frac{1}{\mathcal{Z}_{0}} 
\frac{\partial^{2}\mathcal{Z}^{G} [\mathrm{J}, \mathrm{J}^{*}, \eta, \eta^{*}]}  
{\partial \mathrm{J}_{\bm{k}_{\bm{0}},\sigma_{0},\lambda_{0}} (\mathrm{i} \omega_{m_{0}})
\partial \mathrm{J}_{\bm{k}_{\bm{0}},\sigma_{0},\lambda_{0}}^{*} (\mathrm{i} \omega_{m_{0}})}
\right]_{\begin{array}{l} \mbox{\scriptsize{$\mathrm{J} = \mathrm{J}^{*} = 0$}} \\ 
 \mbox{\scriptsize{$\eta = \eta^{*} = 0$}} \end{array}} 
= - \frac{1}{\mathrm{i} \omega_{m_{0}} - \varepsilon_{\bm{k}_{\bm{0}}}}  \ , \ \\
\label{FreePrFer-Deta2}
& & \left.
\frac{1}{\mathcal{Z}_{0}} 
\frac{\partial^{2}\mathcal{Z}^{G} [\mathrm{J}, \mathrm{J}^{*}, \eta, \eta^{*}]}  
{\partial \eta_{\gamma_{0}} (\mathrm{i} \omega_{m_{0}}) \partial \eta_{\gamma_{0}}^{*}(\mathrm{i} \omega_{m_{0}})}
\right]_{\begin{array}{l} \mbox{\scriptsize{$\mathrm{J} = \mathrm{J}^{*} = 0$}} \\ 
 \mbox{\scriptsize{$\eta = \eta^{*} = 0$}} \end{array}}
= - \frac{\frac{\displaystyle 1}{\displaystyle d_{ \{ \Gamma \} }} \  \sum\limits_{n} \mathrm{e}^{\mathrm{i} \beta \mu_{n}}
\Big(\frac{\displaystyle 1}{\displaystyle \mathrm{i} \omega_{m_{0}} - \mathrm{i}\mu_{n}} \Big)
\mathcal{Z}_{0}^{f} (\mathrm{i}\mu_{n})}
{ \frac{\displaystyle 1}{\displaystyle d_{ \{ \Gamma \} }} \  \sum\limits_{n} \mathrm{e}^{\mathrm{i} \beta \mu_{n}}
 \mathcal{Z}_{0}^{f} (\mathrm{i}\mu_{n}) } \ , \
\end{eqnarray}
\end{widetext}
where $\mathcal{Z}_{0}^{f}$ is the free fermionic part of the partition function, associated with the imaginary chemical potential $(\mathrm{i}\mu_{n})$

\begin{eqnarray}
\label{FreePartF-fermi}
& &\mathcal{Z}_{0}^{f} (\mathrm{i}\mu_{n}) \\
&=&
\int \mathcal{D}\bar{f}^{*} \mathcal{D}\bar{f} 
\mathrm{exp} \Big[ - \sum_{\mathrm{i} \omega_{m}} \sum_{\gamma} \big\{
\bar{f}_{\gamma}^{*}(\mathrm{i} \omega_{m}) (-\mathrm{i} \omega_{m} + \mathrm{i}\mu_{n}) \nonumber \\
& &\hspace{40mm}\times \ \bar{f}_{\gamma}(\mathrm{i} \omega_{m}) \big\} \Big]~. \nonumber
\end{eqnarray}
In Eq.\! (\ref{FreePartF-fermi}) we recover the partition function of free fermions \cite{NegOr} of \textquotedblleft energy\textquotedblright  $(\mathrm{i}\mu_{n})$ 
\begin{eqnarray}
\mathcal{Z}_{0}^{f} (\mathrm{i}\mu_{n}) =
\prod\limits_{\gamma=1}^{d_{ \{ \Gamma \} }} \big[ 1+ \mathrm{e}^{-\mathrm{i} \beta \mu_{n}} \big] =
 \big[ 1+ \mathrm{e}^{-\mathrm{i} \beta \mu_{n}} \big]^{d_{ \{ \Gamma \} }}~.
\label{FreePartF-fermi2}
\end{eqnarray}
Due to the particular form of the imaginary chemical potentials, one can derive some identities with the help of simple trigonometric properties, which will be useful to perform the calculations involving the pseudofermion propagators (see Appendix \ref{Appen-ThirOrd})
 \begin{eqnarray}
\label{Id-SumMu1}
& &\sum _{n=1}^{d_{ \{ \Gamma \} }}  \mathrm{e}^{\mathrm{i} \beta \mu_{n}} 
\big[ 1+ \mathrm{e}^{-\mathrm{i} \beta \mu_{n}} \big]^{d_{ \{ \Gamma \} }} =
\big[d_{ \{ \Gamma \} }\big]^{2} ~, \\
\label{Id-SumMu2}
& &\sum _{n=1}^{d_{ \{ \Gamma \} }} \big[ 1+ \mathrm{e}^{-\mathrm{i} \beta \mu_{n}} \big]^{d_{ \{ \Gamma \} } - 1} =
d_{ \{ \Gamma \} } ~, \\
\label{Id-SumMu3}
& &\sum _{n=1}^{d_{ \{ \Gamma \} }}  \mathrm{e}^{\mathrm{i} \beta \mu_{n}} 
\big[ 1+ \mathrm{e}^{-\mathrm{i} \beta \mu_{n}} \big]^{d_{ \{ \Gamma \} }-1} =
d_{ \{ \Gamma \} } \big(d_{ \{ \Gamma \} } -1\big)~.~~~~~~~~
\end{eqnarray}
 
Using Eqs.\! (\ref{FreePrEl-DJ}) and (\ref{FreePrEl-DJ2}) we get for the free conduction electron propagator
\begin{eqnarray}
\langle \bar{c}_{\bm{k}_{\bm{0}},\sigma_{0},\lambda_{0}}(\mathrm{i} \omega_{m_{0}})
\bar{c}_{\bm{k}_{\bm{0}},\sigma_{0},\lambda_{0}}^{*}(\mathrm{i} \omega_{m_{0}}) \rangle_{0} =
- \frac{1}{\mathrm{i} \omega_{m_{0}} - \varepsilon_{\bm{k}_{\bm{0}}}}~.~~~~~
\label{FreePropEl}
\end{eqnarray}
The expressions (\ref{FreePrFer-Deta}), (\ref{FreePrFer-Deta2}), (\ref{FreePartF-fermi2}) and (\ref{Id-SumMu1}) yield the free propagator associated with the impurity spin pseudofermion
\begin{eqnarray}
\label{FreePropFermi}
& &\langle \bar{f}_{\gamma_{0}}(\mathrm{i} \omega_{m_{0}}) \bar{f}_{\gamma_{0}}^{*}(\mathrm{i} \omega_{m_{0}})  \rangle_{0} \\
&=&
- \frac{1}{\big[d_{ \{ \Gamma \} }\big]^{2}} \  \sum_{n=1}^{d_{ \{ \Gamma \} }} 
\Big(\frac{\displaystyle 1}{\displaystyle \mathrm{i} \omega_{m_{0}} - \mathrm{i}\mu_{n}} \Big)
\mathrm{e}^{\mathrm{i} \beta \mu_{n}} 
\big[ 1+ \mathrm{e}^{-\mathrm{i} \beta \mu_{n}} \big]^{d_{ \{ \Gamma \} }}~. \nonumber
\end{eqnarray}


\section{Computation of the spin factors present in the conduction electron self-energy}
\label{Appen-SpinFact}

In this Appendix, we present the method of calculation of the spin factors necessary to obtain the electronic self-energy perturbatively. The properties of the SU($N$) generators given in Appendix \ref{Appen-SUNgener} will be used intensively. We will detail the contribution of each diagram of Figs.\! \ref{SelfEnDiags-Fig2}, \ref{SelfEnDiags-Fig3} and \ref{SelfEnDiags-Fig4}. A general remark one can make is that the impurity spin propagators form a closed loop in each diagram. Therefore the  $\big\{\tilde{\bm{\Gamma}}^{A} \big\}$ matrices will always appear as a trace over products of generators in the $\{ \Gamma \}$ representation.  


\subsection{Spin multiplicative factor in the second order term of the perturbation calculation of the conduction electron self-energy}
\label{AppenSub-Spin2}

After summation over the internal spin degrees of freedom, the spin factor contribution of the second order term to the electronic self-energy is given by  
\begin{eqnarray}
\mbox{Spin}\Big[ \mbox{\large{$\Sigma$}}_{(2,0)}^{\sigma, \sigma^{'}} \Big] = 
\sum_{A,B=1}^{N^{2}-1} \mathrm{Tr}\big[ \tilde{\bm{\Gamma}}^{A} \tilde{\bm{\Gamma}}^{B} \big] \cdot
\big[ \tilde{\bm{\sigma}}^{B} \tilde{\bm{\sigma}}^{A}\big]_{\sigma^{'} \sigma}~.
\label{Def-Spin2}
\end{eqnarray}

The SU($N$) generators in the $\{ \Gamma \}$ representation obey the following normalization relation \cite{Okubo-1982}
\begin{eqnarray}
 \mathrm{Tr}\big[ \tilde{\bm{\Gamma}}^{A} \tilde{\bm{\Gamma}}^{B} \big] =
\frac{d_{ \{ \Gamma \} } \mathcal{C}_{2}\big(\{ \Gamma \} \big)}{d_{ \{ \sigma \} } \mathcal{C}_{2}\big(\{ \sigma \} \big)} 
\mathrm{Tr}\big[ \tilde{\bm{\sigma}}^{A} \tilde{\bm{\sigma}}^{B} \big]~,
\label{Norm-Gamma2} 
\end{eqnarray}
which gives by making use of Eqs.\! (\ref{Dim-sig}), (\ref{Casi-sig}), (\ref{Norm-sig}), (\ref{Dim-Gam}) and (\ref{Casi-Gam-full})
\begin{eqnarray}
\label{Tr-Spin2}
& &\mathrm{Tr}\big[ \tilde{\bm{\Gamma}}^{A} \tilde{\bm{\Gamma}}^{B} \big] \\
&=& \frac{(N+2S-1)!}{(2S)! (q-1)! (N-q)!}  
\cdot \frac{2S}{2N (N^{2}-1)} \nonumber \\
& & \times \ \big[ 2S(N-1) - q(N+1) + (N^{2}+1) \big] \delta_{AB}~, \nonumber
\end{eqnarray}
where $2S$ and $q$ are the parameters of the Young tableau associated with the L-shaped representation, as shown in Fig.\! \ref{LShape-Fig}.

Using Eqs.\! (\ref{Tr-Spin2}) and (\ref{Compl-sig}) we obtain finally the spin multiplicative factor of the second order term  
\begin{eqnarray}
\label{Contrib-Spin2}
& &\mbox{Spin}\Big[ \mbox{\large{$\Sigma$}}_{(2,0)}^{\sigma, \sigma^{'}} \Big] \\
&=& \frac{(N+2S-1)!}{(2S)! (q-1)! (N-q)!}  \cdot \frac{2S}{4N^{2}} \nonumber \\
& & \times \ \Big[ 2S(N-1) - q(N+1) + (N^{2}+1) \Big] \delta_{\sigma, \sigma^{'}}~. \nonumber
\end{eqnarray}


\subsection{Spin multiplicative factors in the third order term of the perturbation calculation of the conduction electron self-energy}
\label{AppenSub-Spin3}

After summation over the internal spin degrees of freedom, the first spin contribution of the third order term is given by the diagram corresponding to $\mbox{\large{$\Sigma$}}_{(3,0)\mbox{\scriptsize{a}}}^{\sigma, \sigma^{'}}$ 
\begin{eqnarray}
\label{Def-Spin3}
& &\mbox{Spin}\Big[ \mbox{\large{$\Sigma$}}_{(3,0)\mbox{\scriptsize{a}}}^{\sigma, \sigma^{'}} \Big] \\
&=& \sum_{A,B,C=1}^{N^{2}-1} 
\mathrm{Tr}\big[ \tilde{\bm{\Gamma}}^{A} \tilde{\bm{\Gamma}}^{C} \tilde{\bm{\Gamma}}^{B} \big] \cdot
\big[ \tilde{\bm{\sigma}}^{B} \tilde{\bm{\sigma}}^{C} \tilde{\bm{\sigma}}^{A}\big]_{\sigma^{'} \sigma}~. \nonumber
\end{eqnarray}

The calculation of $\mathrm{Tr}\big[ \tilde{\bm{\Gamma}}^{A} \tilde{\bm{\Gamma}}^{C} \tilde{\bm{\Gamma}}^{B} \big]$ for the L-shaped representation in the SU($N$) group is rather technical. Fortunately this kind of algebraic object has already been intensively studied in the field of particle physics. We can express it with the aid of a symmetrized trace, which defines a symmetric invariant tensor associated with $\{ \Gamma \}$
\cite{Schell-1999}
\begin{eqnarray}
\mathrm{Tr}\big[ \tilde{\bm{\Gamma}}^{A} \tilde{\bm{\Gamma}}^{C} \tilde{\bm{\Gamma}}^{B} \big] =
\mathrm{STr}_{ \{ \Gamma \} }^{ACB} +
\frac{\mathrm{i}}{2} \frac{d_{ \{ \Gamma \} } \mathcal{C}_{2}\big(\{ \Gamma \} \big)}{N^{2}-1}f^{ACB}~, ~~
\label{Def-TrGam3}
\end{eqnarray}
where $\mathrm{STr}_{ \{ \Gamma \} }^{ACB}$ is the third order symmetrized trace
\begin{eqnarray}
\mathrm{STr}_{ \{ \Gamma \} }^{A_{1} A_{2} A_{3} } = 
\frac{1}{3!} \sum_{\pi}  \mathrm{Tr}\big[
\tilde{\bm{\Gamma}}^{A_{\pi(1)}} \tilde{\bm{\Gamma}}^{A_{\pi(2)}} \tilde{\bm{\Gamma}}^{A_{\pi(3)}}\big]~. \nonumber
\label{Def-STr}
\end{eqnarray}
In the previous expression the summation is performed over all the permutations of the generator indices. $\mathrm{STr}_{ \{ \Gamma \} }^{ACB}$ can be calculated for any representation in SU($N$) ($N \geq 3$) by following the method developed by Okubo, \cite{Okubo-1977-2} that we summarize here. The Young tableau in the standard arrangement of the L-shaped representation is specified by $N$ integers $(f_{i})_{1 \leq i \leq N}$, $f_{i}$ being the number of boxe(s) in the $i$th row
\begin{eqnarray}
\left\lbrace
\begin{array}{l}
f_{1} = 2S \\
f_{i} = 1 \ , \ 2 \leq i \leq q \\
f_{i} = 0 \ , \ q+1 \leq i \leq N. \nonumber
\end{array}
\right.
\label{Def-find}
\end{eqnarray}
We define the indices $(\sigma_{j})_{1 \leq j \leq N}$ by
\begin{eqnarray}
\sigma_{j} &=& f_{j} + \frac{N+1}{2} - j - \frac{1}{N} \sum_{i=1}^{N} f_{i} \nonumber \\
&=&
\frac{N+1}{2} - \frac{2S+q-1}{N} - (j- f_{j}),  \nonumber
\label{Def-sigind}
\end{eqnarray}
then
\begin{eqnarray}
\mathrm{STr}_{ \{ \Gamma \} }^{ACB} = \frac{d^{ACB}}{2} K\big(\{ \Gamma \}\big)~,
\label{STr3-interm}
\end{eqnarray}
with
\begin{eqnarray}
K\big(\{ \Gamma \}\big) =
\frac{N}{(N^{2}-1)(N^{2}-4)} d_{ \{ \Gamma \} } \sum_{j=1}^{N} (\sigma_{j})^{3}~.
\label{Def-KTr3}
\end{eqnarray}
After lengthy algebra we get using Eqs.\! (\ref{Dim-Gam}), (\ref{Casi-Gam-full}), (\ref{Def-TrGam3}), (\ref{STr3-interm}) and (\ref{Def-KTr3})
\begin{widetext}
\begin{eqnarray}
\mathrm{Tr}\big[ \tilde{\bm{\Gamma}}^{A} \tilde{\bm{\Gamma}}^{C} \tilde{\bm{\Gamma}}^{B} \big] &=&
\frac{(N+2S-1)!}{(2S)! (q-1)! (N-q)!} \cdot \frac{2S}{2 (N^{2}-1)} 
\nonumber \\
& &\times \ 
\Bigg\{ \frac{1}{(N^{2}-4)} \Bigg((2S)^{2} \Bigg[ \frac{2}{N} - 3 + N \Bigg] + q^{2} \Bigg[\frac{2}{N} + 3 + N  \Bigg] +
2S \cdot q \Bigg[\frac{4}{N} - N \Bigg] \nonumber \\
& & \hspace{22mm} +2S \Bigg[-\frac{4}{N} + 3 -\frac{7N}{2} +\frac{3N^{2}}{2} \Bigg] +
q \Bigg[-\frac{4}{N} - 3 -\frac{7N}{2} -\frac{3N^{2}}{2} \Bigg] 
\nonumber \\
& &\hspace{22mm} 
+ \Bigg[\frac{2}{N}+ \frac{7N}{2} + \frac{N^{3}}{2}\Bigg] \Bigg) d^{ACB} \nonumber \\
& & \hspace{5mm}+ \frac{\mathrm{i}}{2} \Bigg(2S \Bigg[1-\frac{1}{N}\Bigg] - q \Bigg[1+\frac{1}{N}\Bigg]+ \Bigg[N+\frac{1}{N}\Bigg]
\Bigg)f^{ACB} \Bigg\} ~. 
\label{TrGam3-fin}
\end{eqnarray}
\end{widetext}
We have checked the validity of the previous expression (\ref{TrGam3-fin}) by comparing it with the results of Okubo and Patera \cite{OkuPat-1983} who considered special L-shaped representations, i.e., with given value of $2S$ and $q$ for general SU($N$) groups. We get a perfect agreement.

For the electronic part we can use
\begin{eqnarray}
\label{ElecPart3}
\big[ \tilde{\bm{\sigma}}^{B} \tilde{\bm{\sigma}}^{C} \tilde{\bm{\sigma}}^{A}\big]_{\sigma^{'} \sigma} 
&=& \frac{1}{2N}\delta_{BC} \big[\tilde{\bm{\sigma}}^{A}\big]_{\sigma^{'} \sigma}  \\
&+& \frac{1}{2}\sum_{G=1}^{N^{2}-1}\big(d^{BCG} + \mathrm{i}f^{BCG} \big) 
\big[\tilde{\bm{\sigma}}^{G} \tilde{\bm{\sigma}}^{A} \big]_{\sigma^{'} \sigma}~. \nonumber
\end{eqnarray}

By combining Eqs.\! (\ref{Compl-sig}), (\ref{Multdd}), (\ref{Multff}), (\ref{Multdf}) and the previous results (\ref{TrGam3-fin}),
(\ref{ElecPart3}), we obtain 
\begin{eqnarray}
\label{Contrib-Spin3}
& &\mbox{Spin}\Big[ \mbox{\large{$\Sigma$}}_{(3,0)\mbox{\scriptsize{a}}}^{\sigma, \sigma^{'}} \Big] \\
&=&
\frac{(N+2S-1)!}{(2S)! (q-1)! (N-q)!} \cdot \frac{2S}{8 N^{3}} \nonumber \\
& &\times \ \Big[\hspace{2mm} (2S)^{2} \big(2-3N+N^{2} \big) +  q^{2} \big(2+3N+N^{2} \big) \nonumber \\
& & \hspace{5mm}+ 2S \cdot q  \big(4 - N^{2} \big) 
+ 2S \big(-4+3N -4N^{2} + 2N^{3}\big) \nonumber \\
& & \hspace{5mm}+ q  \big(-4-3N -4N^{2} - 2N^{3}\big) \nonumber \\
& & \hspace{5mm}+ \big(2+4N^{2}+N^{4}\big) \Big] \delta_{\sigma, \sigma^{'}} ~. \nonumber
\end{eqnarray}

The second spin contribution of the third order term corresponds to the diagram $\mbox{\large{$\Sigma$}}_{(3,0)\mbox{\scriptsize{b}}}^{\sigma, \sigma^{'}}$ 
\begin{eqnarray}
\label{Def-Spin3b}
& &\mbox{Spin}\Big[ \mbox{\large{$\Sigma$}}_{(3,0)\mbox{\scriptsize{b}}}^{\sigma, \sigma^{'}} \Big] \\
&=& \sum_{A,B,C=1}^{N^{2}-1} 
\mathrm{Tr}\big[ \tilde{\bm{\Gamma}}^{A} \tilde{\bm{\Gamma}}^{C} \tilde{\bm{\Gamma}}^{B} \big] \cdot
\big[ \tilde{\bm{\sigma}}^{C} \tilde{\bm{\sigma}}^{B} \tilde{\bm{\sigma}}^{A}\big]_{\sigma^{'} \sigma}~, \nonumber
\end{eqnarray}
which can be derived in a similar way. The result is
\begin{eqnarray}
\label{Contrib-Spin3b}
& &\mbox{Spin}\Big[ \mbox{\large{$\Sigma$}}_{(3,0)\mbox{\scriptsize{b}}}^{\sigma, \sigma^{'}} \Big] \\
&=& 
\frac{(N+2S-1)!}{(2S)! (q-1)! (N-q)!} \cdot \frac{2S}{8 N^{3}} \nonumber \\
& &\times \Big[\hspace{2mm} (2S)^{2} \big(2-3N+N^{2} \big) +  q^{2} \big(2+3N+N^{2} \big) \nonumber \\
& & \hspace{5mm} + 2S \cdot q  \big(4 - N^{2} \big) 
+ 2S \big(-4+3N -3N^{2} + N^{3}\big) \nonumber \\
& & \hspace{5mm}+ q  \big(-4-3N -3N^{2} - N^{3}\big) 
+ \big(2+3N^{2}\big) \Big] \delta_{\sigma, \sigma^{'}} ~. \nonumber
\end{eqnarray}


\subsection{Spin multiplicative factors in the fourth order term of the perturbation calculation of the conduction electron self-energy}
\label{AppenSub-Spin4}

The first spin contribution resulting from the fourth order term is given by the diagram corresponding to
$\mbox{\large{$\Sigma$}}_{(4,1)\mbox{\scriptsize{a}}}^{\sigma, \sigma^{'}}$ 
\begin{eqnarray}
\label{Def-Spin41a}
& &\mbox{Spin}\Big[ \mbox{\large{$\Sigma$}}_{(4,1)\mbox{\scriptsize{a}}}^{\sigma, \sigma^{'}} \Big] \\
&=& 
\sum_{A,B,C,D=1}^{N^{2}-1} 
\mathrm{Tr}\big[\tilde{\bm{\Gamma}}^{D} \tilde{\bm{\Gamma}}^{C} \tilde{\bm{\Gamma}}^{B} \tilde{\bm{\Gamma}}^{A}  \big] \cdot
\mathrm{Tr}\big[ \tilde{\bm{\sigma}}^{B} \tilde{\bm{\sigma}}^{C}  \big] \nonumber \\
& &\hspace{14mm} \times \ \big[ \tilde{\bm{\sigma}}^{D} \tilde{\bm{\sigma}}^{A}\big]_{\sigma^{'} \sigma}~. \nonumber
\end{eqnarray}
With (\ref{Norm-sig}) and (\ref{ProdGam-4}) we get
\begin{eqnarray}
\label{Def-TrGam41a}
& &\sum_{B,C}
\mathrm{Tr}\big[\tilde{\bm{\Gamma}}^{D} \tilde{\bm{\Gamma}}^{C} \tilde{\bm{\Gamma}}^{B} \tilde{\bm{\Gamma}}^{A}  \big] \cdot
\mathrm{Tr}\big[ \tilde{\bm{\sigma}}^{B} \tilde{\bm{\sigma}}^{C}  \big] \\
&=&
\frac{(N+2S-1)!}{(2S)! (q-1)! (N-q)!} \cdot \frac{2S \big(2S+q-1 \big)}{8 N^{2} \big(N^{2} -1\big)} \nonumber \\
& &\times  \Big[2S \big(N-1\big) -q\big(N+1\big) +\big(N^{2}+1\big) \Big]^{2} \delta_{AD}~. \nonumber
\end{eqnarray}
Using Eq.\! (\ref{Mult-law}) we have
\begin{eqnarray}
\label{Contrib-Spin41a}
& &\mbox{Spin}\Big[ \mbox{\large{$\Sigma$}}_{(4,1)\mbox{\scriptsize{a}}}^{\sigma, \sigma^{'}} \Big] \\
&=&
\frac{(N+2S-1)!}{(2S)! (q-1)! (N-q)!} \cdot \frac{2S \big(2S+q-1 \big)}{16 N^{3}} \nonumber \\
& &\times  \Big[2S \big(N-1\big) -q\big(N+1\big) +\big(N^{2}+1\big) \Big]^{2} \delta_{\sigma, \sigma^{'}} ~. \nonumber
\end{eqnarray}
 
The second spin contribution resulting from the fourth order diagram 
$\mbox{\large{$\Sigma$}}_{(4,1)\mbox{\scriptsize{b}}}^{\sigma, \sigma^{'}}$ is
\begin{eqnarray}
\mbox{Spin}\Big[ \mbox{\large{$\Sigma$}}_{(4,1)\mbox{\scriptsize{b}}}^{\sigma, \sigma^{'}} \Big] 
&=&
\sum_{A,B,C,D=1}^{N^{2}-1} 
\mathrm{Tr}\big[ \tilde{\bm{\Gamma}}^{A} \tilde{\bm{\Gamma}}^{B} \big] \cdot
\mathrm{Tr}\big[ \tilde{\bm{\Gamma}}^{C} \tilde{\bm{\Gamma}}^{D} \big] \nonumber ~~~~~~~ \\
& &\hspace{6mm}\times \ \mathrm{Tr}\big[ \tilde{\bm{\sigma}}^{B} \tilde{\bm{\sigma}}^{C}  \big] \cdot
\big[ \tilde{\bm{\sigma}}^{D} \tilde{\bm{\sigma}}^{A}\big]_{\sigma^{'} \sigma}~.
\label{Def-Spin41b}
\end{eqnarray}
By using Eqs.\! (\ref{Norm-sig}), (\ref{Compl-sig}) and (\ref{Norm-Gam}) we get
\begin{eqnarray}
\label{Contrib-Spin41b}
& &\mbox{Spin}\Big[ \mbox{\large{$\Sigma$}}_{(4,1)\mbox{\scriptsize{b}}}^{\sigma, \sigma^{'}} \Big] \\
&=& 
\Bigg\{ \frac{(N+2S-1)!}{(2S)! (q-1)! (N-q)!} \cdot \frac{2S}{N}  \nonumber \\
& &\hspace{4mm} \times \Big[2S \big(N-1\big) -q\big(N+1\big) +\big(N^{2}+1\big) \Big] \Bigg\}^{2} \nonumber \\
& &\times  \frac{1}{16 N \big(N^{2}-1\big)} \ \delta_{\sigma, \sigma^{'}} ~. \nonumber
\end{eqnarray}

The last spin multiplicative factor showing up in the fourth order diagram 
$\mbox{\large{$\Sigma$}}_{(4,1)\mbox{\scriptsize{c}}}^{\sigma, \sigma^{'}}$ is given by
\begin{eqnarray}
\label{Def-Spin41c}
\mbox{Spin}\Big[ \mbox{\large{$\Sigma$}}_{(4,1)\mbox{\scriptsize{c}}}^{\sigma, \sigma^{'}} \Big] &=& 
\sum_{A,B,C,D=1}^{N^{2}-1} 
\mathrm{Tr}\big[\tilde{\bm{\Gamma}}^{A} \tilde{\bm{\Gamma}}^{C} \tilde{\bm{\Gamma}}^{D} \tilde{\bm{\Gamma}}^{B}  \big] 
  \\
& &\hspace{12mm} \times \ \mathrm{Tr}\big[ \tilde{\bm{\sigma}}^{B} \tilde{\bm{\sigma}}^{C}  \big] \cdot
\big[ \tilde{\bm{\sigma}}^{D} \tilde{\bm{\sigma}}^{A}\big]_{\sigma^{'} \sigma}~. \nonumber
\end{eqnarray}
The expressions (\ref{Norm-sig}) and (\ref{ProdGam-3}) lead to
\begin{eqnarray}
\label{Def-TrGam41c}
& &\sum_{B,C}
\mathrm{Tr}\big[\tilde{\bm{\Gamma}}^{A} \tilde{\bm{\Gamma}}^{C} \tilde{\bm{\Gamma}}^{D} \tilde{\bm{\Gamma}}^{B}  \big] \cdot
\mathrm{Tr}\big[ \tilde{\bm{\sigma}}^{B} \tilde{\bm{\sigma}}^{C}  \big] \\
&=& \frac{(N+2S-1)!}{(2S)! (q-1)! (N-q)!} \cdot \frac{2S}{4N \big(N^{2}-1\big)}  \nonumber \\
& &\times  \Bigg[ \frac{2S+q-1}{2N} \Big\{2S \big(N-1\big) -q\big(N+1\big) \nonumber \\
& &\hspace{24mm} +\big(N^{2}+1\big) \Big\} 
- \frac{N}{2} \Bigg]
 \nonumber \\
&  &\times \hspace{2mm}
 \Big[2S \big(N-1\big) -q\big(N+1\big) +\big(N^{2}+1\big) \Big] \delta_{AD}~. \nonumber
\end{eqnarray}

Using Eq.\! (\ref{Mult-law}) we obtain 
\begin{eqnarray}
\label{Contrib-Spin41c}
& &\mbox{Spin}\Big[ \mbox{\large{$\Sigma$}}_{(4,1)\mbox{\scriptsize{c}}}^{\sigma, \sigma^{'}} \Big] \\
&=&
\frac{(N+2S-1)!}{(2S)! (q-1)! (N-q)!} \cdot \frac{2S}{16 N^{3}}  \cdot \delta_{\sigma, \sigma^{'}} \nonumber \\
& &\times \ \Big\{ \hspace{2mm} \Big[2S \big(N-1\big) -q\big(N+1\big) +\big(N^{2}+1\big) \Big]^{2} \nonumber \\
& & \hspace{7mm} \times \ (2S+q-1) \nonumber \\
& & \hspace{5mm} - \Big[2S \big(N-1\big) -q\big(N+1\big) +\big(N^{2}+1\big) \Big] N^{2} \Big\} ~. \nonumber
\end{eqnarray}


\section{Explicit calculation of a third order perturbative diagram}
\label{Appen-ThirOrd}

In this Appendix we show how to compute the contribution due to the diagram associated with $\mbox{\large{$\Sigma$}}_{(3,0)\mbox{\scriptsize{a}}}^{\sigma, \sigma^{'}}$ in  Fig.\! \ref{SelfEnDiags-Fig3}. All the other diagrams can be treated in the same way. The spin multiplicative factor resulting from the $\big\{ \tilde{\bm{\Gamma}}^{A} \big\}$ and $\big\{ \tilde{\bm{\sigma}}^{A} \big\}$ matrices has been calculated in the previous Section \ref{AppenSub-Spin3}, it is given by Eq.\! (\ref{Contrib-Spin3}). Here we just show how its starting point (\ref{Def-Spin3}) can be derived. Then we will focus on the frequency and momentum part. In  Fig.\! \ref{3OrdDiag-Fig} the diagram associated with $\mbox{\large{$\Sigma$}}_{(3,0)a}^{\sigma, \sigma^{'}}$ is represented, by including all the internal degrees of freedom.

  Let us start with the spin contribution. According to the rules defined in Section \ref{Sub-ConstDiag}, we obtain by expliciting the diagram shown in  Fig.\! \ref{3OrdDiag-Fig}
\begin{figure*}[t]
\includegraphics[width=10.0cm]{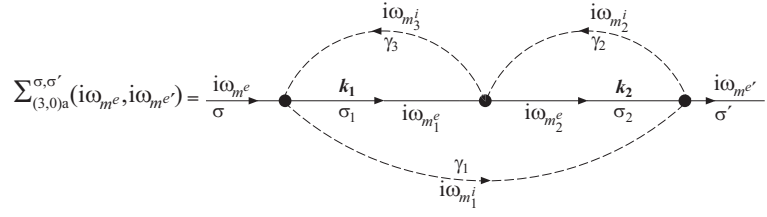}
\caption {Feynman diagram of the third order in perturbation for the conduction electron self-energy (according to the notations defined in Section \ref{Sub-ConstDiag}).} 
\label{3OrdDiag-Fig}
\end{figure*}
\begin{eqnarray}
& & \mbox{Spin}\Big[ \mbox{\large{$\Sigma$}}_{(3,0)\mbox{\scriptsize{a}}}^{\sigma, \sigma^{'}} \Big] \nonumber \\
&=&
\sum _{\sigma_{1},\sigma_{2}=1}^{N} \sum _{\gamma_{1},\gamma_{2},\gamma_{3}=1}^{d_{ \{ \Gamma \} }}
\Big( \sum_{A=1}^{N^{2}-1} \bm{\sigma}_{\sigma_{1},\sigma}^{A} \bm{\Gamma}_{\gamma_{1},\gamma_{3}}^{A} \Big) \nonumber \\
& & \hspace{24mm} \times \ 
\Big( \sum_{B=1}^{N^{2}-1} \bm{\sigma}_{\sigma^{'},\sigma_{2}}^{B} \bm{\Gamma}_{\gamma_{2},\gamma_{1}}^{B} \Big) \nonumber \\
& & \hspace{24mm} \times \ 
\Big( \sum_{C=1}^{N^{2}-1} \bm{\sigma}_{\sigma_{2},\sigma_{1}}^{C} \bm{\Gamma}_{\gamma_{3},\gamma_{2}}^{C} \Big) \nonumber \\ 
&=& \sum_{A,B,C} \big[ \tilde{\bm{\sigma}}^{B} \tilde{\bm{\sigma}}^{C} \tilde{\bm{\sigma}}^{A}\big]_{\sigma^{'} \sigma} \cdot
\mathrm{Tr}\big[ \tilde{\bm{\Gamma}}^{A} \tilde{\bm{\Gamma}}^{C} \tilde{\bm{\Gamma}}^{B} \big]~,
\label{DefDev-Spin3}
\end{eqnarray}
which is identical to Eq.\! (\ref{Def-Spin3}).

  We consider now the contribution including the frequencies and the momentum, denoted by $F_{(3,0)\mbox{\scriptsize{a}}}$. Although our analysis is restricted to the zero-temperature limit, the calculations are performed in the space of (fermionic) Matsubara frequencies for reasons of simplicity. We shall use the properties of the free propagators presented in Appendix \ref{Appen-FreeProp}. The rules defined in Section \ref{Sub-ConstDiag} combined with expressions (\ref{FreePropEl}) and  (\ref{FreePropFermi}) give
\begin{widetext}
\begin{eqnarray}
F_{(3,0)\mbox{\scriptsize{a}}} = & &\left(- \frac{J_{\mathrm{K}}}{\mathcal{N}_{s}}\right)^{3} \sum_{\bm{k}_{\bm{1}},\bm{k}_{\bm{2}}} \left(\frac{1}{\beta} \right)^{3}
\sum_{m_{1}^{e},m_{2}^{e}} \sum_{m_{1}^{i},m_{2}^{i},m_{3}^{i}} 
\delta_{m^{e}+m_{3}^{i},m_{1}^{e}+m_{1}^{i}} \cdot \delta_{m_{1}^{e}+m_{2}^{i},m_{2}^{e}+m_{3}^{i}} \cdot 
\delta_{m_{2}^{e}+m_{1}^{i},m^{e'}+m_{2}^{i}} \nonumber \\
&\times& \frac{(-1)}{\big(\mathrm{i} \omega_{m_{1}^{e}} - \varepsilon_{\bm{k}_{\bm{1}}}\big)} \cdot \frac{(-1)}{\big(\mathrm{i} \omega_{m_{2}^{e}} - \varepsilon_{\bm{k}_{\bm{2}}}\big)}
\cdot \left(- \frac{1}{\big[d_{ \{ \Gamma \} }\big]^{2}} \  \sum_{n_{1}} 
\frac{\mathrm{e}^{\mathrm{i} \beta \mu_{n_{1}}} \big[ 1+ \mathrm{e}^{-\mathrm{i} \beta \mu_{n_{1}}} \big]^{d_{ \{ \Gamma \} }}}
{\mathrm{i} \omega_{m_{1}^{i}} - \mathrm{i}\mu_{n_{1}}} \right) 
\nonumber \\ 
&\times& 
\left(- \frac{1}{\big[d_{ \{ \Gamma \} }\big]^{2}} \  \sum_{n_{2}} 
\frac{\mathrm{e}^{\mathrm{i} \beta \mu_{n_{2}}} \big[ 1+ \mathrm{e}^{-\mathrm{i} \beta \mu_{n_{2}}} \big]^{d_{ \{ \Gamma \} }}}
{\mathrm{i} \omega_{m_{2}^{i}} - \mathrm{i}\mu_{n_{2}}} \right) 
\cdot \left(- \frac{1}{\big[d_{ \{ \Gamma \} }\big]^{2}} \  \sum_{n_{3}} 
\frac{\mathrm{e}^{\mathrm{i} \beta \mu_{n_{3}}} \big[ 1+ \mathrm{e}^{-\mathrm{i} \beta \mu_{n_{3}}} \big]^{d_{ \{ \Gamma \} }}}
{\mathrm{i} \omega_{m_{3}^{i}} - \mathrm{i}\mu_{n_{3}}} \right)~.
\label{Def-F30}
\end{eqnarray}
After summation over the imaginary frequencies we get

\begin{eqnarray}
F_{(3,0)\mbox{\scriptsize{a}}} = & & \left(- \frac{J_{\mathrm{K}}}{\mathcal{N}_{s}}\right)^{3} \sum_{\bm{k}_{\bm{1}},\bm{k}_{\bm{2}}} \delta_{m^{e},m^{e'}}
\left(  \frac{1}{\big[d_{ \{ \Gamma \} }\big]^{2}} \  \sum_{n_{1}} 
\mathrm{e}^{\mathrm{i} \beta \mu_{n_{1}}} \big[ 1+ \mathrm{e}^{-\mathrm{i} \beta \mu_{n_{1}}} \big]^{d_{ \{ \Gamma \} }} \right) \nonumber \\ 
&\times& \left(  \frac{1}{\big[d_{ \{ \Gamma \} }\big]^{2}} \  \sum_{n_{2}} 
\mathrm{e}^{\mathrm{i} \beta \mu_{n_{2}}} \big[ 1+ \mathrm{e}^{-\mathrm{i} \beta \mu_{n_{2}}} \big]^{d_{ \{ \Gamma \} }} \right) \cdot
\left(  \frac{1}{\big[d_{ \{ \Gamma \} }\big]^{2}} \  \sum_{n_{3}} 
\mathrm{e}^{\mathrm{i} \beta \mu_{n_{3}}} \big[ 1+ \mathrm{e}^{-\mathrm{i} \beta \mu_{n_{3}}} \big]^{d_{ \{ \Gamma \} }} \right) \nonumber \\ 
&\times& \big[ n_{\mathrm{FD}}(\varepsilon_{\bm{k}_{\bm{1}}}) - n_{\mathrm{FD}}(\mathrm{i} \mu_{n_{3}} )\big] \cdot
\big[ n_{\mathrm{FD}}(\varepsilon_{\bm{k}_{\bm{2}}}) - n_{\mathrm{FD}}(\mathrm{i}  \mu_{n_{2}} )\big] \nonumber \\ 
&\times& \Bigg[ \frac{n_{\mathrm{FD}}(\mathrm{i} \omega_{m^{e}} -  \varepsilon_{\bm{k}_{\bm{1}}} + \mathrm{i} \mu_{n_{3}}) - n_{\mathrm{FD}}(\mathrm{i} \mu_{n_{1}} )}
{(\mathrm{i} \omega_{m^{e}} -  \varepsilon_{\bm{k}_{\bm{1}}} - \mathrm{i}  \mu_{n_{1}} + \mathrm{i} \mu_{n_{3}} )
(\varepsilon_{\bm{k}_{\bm{1}}} - \varepsilon_{\bm{k}_{\bm{2}}} + \mathrm{i} \mu_{n_{2}} - \mathrm{i} \mu_{n_{3}}) } 
\nonumber \\ 
& & 
+ \frac{n_{\mathrm{FD}}(\mathrm{i} \omega_{m^{e}} -  \varepsilon_{\bm{k}_{\bm{2}}} + \mathrm{i} \mu_{n_{2}}) - 
n_{\mathrm{FD}}(\mathrm{i} \mu_{n_{1}} )}
{(\mathrm{i} \omega_{m^{e}} -  \varepsilon_{\bm{k}_{\bm{2}}} - \mathrm{i}  \mu_{n_{1}} + \mathrm{i} \mu_{n_{2}} )
(-\varepsilon_{\bm{k}_{\bm{1}}} + \varepsilon_{\bm{k}_{\bm{2}}} - \mathrm{i} \mu_{n_{2}} + \mathrm{i} \mu_{n_{3}} )} \Bigg]~,
\label{F30-pole}
\end{eqnarray}
where $n_{\mathrm{FD}}$ is the Fermi-Dirac distribution function. We see in Eq.\! (\ref{F30-pole}) that the poles of the free propagators are not on the real axis, as usual. Resulting from the handling of the fermionic constraint (\ref{Def-Pmu}), the poles are shifted by the imaginary chemical potentials $\mathrm{i} \mu_{n}$. By using the following contraction identity
\begin{eqnarray}
& & \big[ n_{\mathrm{FD}}(\varepsilon_{\bm{k}_{\bm{1}}}) - n_{\mathrm{FD}}(\mathrm{i} \mu_{n_{3}} )\big] \cdot
\big[ n_{\mathrm{FD}}(\mathrm{i} \omega_{m^{e}} -  \varepsilon_{\bm{k}_{\bm{1}}} + \mathrm{i} \mu_{n_{3}}) -
 n_{\mathrm{FD}}(\mathrm{i} \mu_{n_{1}} ) \big] \nonumber \\ 
&=&
- n_{\mathrm{FD}}(-\varepsilon_{\bm{k}_{\bm{1}}}) n_{\mathrm{FD}}(-\mathrm{i} \mu_{n_{1}} ) n_{\mathrm{FD}}(\mathrm{i} \mu_{n_{3}} )
- n_{\mathrm{FD}}(\varepsilon_{\bm{k}_{\bm{1}}}) n_{\mathrm{FD}}(\mathrm{i} \mu_{n_{1}} ) n_{\mathrm{FD}}(-\mathrm{i} \mu_{n_{3}} )~,
\label{ImPot-contract}
\end{eqnarray}
we have
\begin{eqnarray}
F_{(3,0)\mbox{\scriptsize{a}}} = & & 2  \left(\frac{J_{\mathrm{K}}}{\mathcal{N}_{s}}\right)^{3} \sum_{\bm{k}_{\bm{1}},\bm{k}_{\bm{2}}} \delta_{m^{e},m^{e'}}
\left(  \frac{1}{\big[d_{ \{ \Gamma \} }\big]^{2}} \  \sum_{n_{1}} 
\mathrm{e}^{\mathrm{i} \beta \mu_{n_{1}}} \big[ 1+ \mathrm{e}^{-\mathrm{i} \beta \mu_{n_{1}}} \big]^{d_{ \{ \Gamma \} }} \right) \nonumber \\ 
&\times& \left(  \frac{1}{\big[d_{ \{ \Gamma \} }\big]^{2}} \  \sum_{n_{2}} 
\mathrm{e}^{\mathrm{i} \beta \mu_{n_{2}}} \big[ 1+ \mathrm{e}^{-\mathrm{i} \beta \mu_{n_{2}}} \big]^{d_{ \{ \Gamma \} }} \right) \cdot
\left(  \frac{1}{\big[d_{ \{ \Gamma \} }\big]^{2}} \  \sum_{n_{3}} 
\mathrm{e}^{\mathrm{i} \beta \mu_{n_{3}}} \big[ 1+ \mathrm{e}^{-\mathrm{i} \beta \mu_{n_{3}}} \big]^{d_{ \{ \Gamma \} }} \right) \nonumber \\ 
&\times& \big[ n_{\mathrm{FD}}(\varepsilon_{\bm{k}_{\bm{2}}}) - n_{\mathrm{FD}}(\mathrm{i}  \mu_{n_{2}} )\big] 
\cdot \Bigg[ \frac{  n_{\mathrm{FD}}(-\varepsilon_{\bm{k}_{\bm{1}}}) n_{\mathrm{FD}}(-\mathrm{i} \mu_{n_{1}} ) 
n_{\mathrm{FD}}(\mathrm{i} \mu_{n_{3}} )
+ n_{\mathrm{FD}}(\varepsilon_{\bm{k}_{\bm{1}}}) n_{\mathrm{FD}}(\mathrm{i} \mu_{n_{1}} ) n_{\mathrm{FD}}(-\mathrm{i} \mu_{n_{3}} )}
{(\mathrm{i} \omega_{m^{e}} -  \varepsilon_{\bm{k}_{\bm{1}}} - \mathrm{i}  \mu_{n_{1}} + \mathrm{i} \mu_{n_{3}} )
(\varepsilon_{\bm{k}_{\bm{1}}} - \varepsilon_{\bm{k}_{\bm{2}}} + \mathrm{i} \mu_{n_{2}} - \mathrm{i} \mu_{n_{3}}) } \Bigg]~.~~~~~~
\label{F30-interm}
\end{eqnarray}
\end{widetext}
By examining the previous expression (\ref{F30-interm}) one can suspect a technical difficulty in perfor\-ming the summations over the imaginary chemical potentials (labelled as $n_{1}$, $n_{2}$ and $n_{3}$), due to their presence in the denominator. However a simplification occurs, because our study is restricted to the zero-temperature limit. Indeed the auxiliary chemical potentials are proportional to the temperature. Therefore their contribution to the denominator of Eq.\! (\ref{F30-interm}) cancels out at zero-temperature. Of course such a simplification would not be valid if it was applied before performing all the summations over the Matsubara frequencies: it would simply mean that the fermionic constraint (\ref{Cons-fermi}) is neglected, i.e., the considered Hilbert space includes unphysical fermionic states. However once the summations over the Matsubara frequencies have been computed, all the poles of the free propagators have been taken into account. Consequently the elimination of the imaginary chemical potentials from the denominator at this stage does not suppress any ``hidden'' property and is perfectly justified. We have to notice also in Eq.\! (\ref{F30-interm}) that the contributions of the imaginary chemical potentials remain at zero-temperature in the exponential factors, contrary to what happens in the denominator. Indeed the terms $\mathrm{e}^{\mathrm{i} \beta \mu_{n}}$ are purely trigonometric factors, independent of the temperature. It is probably feasible to extend the study to the finite temperature limit, for example by performing the summations over the imaginary chemical potentials numerically. This generalization is left for further studies.

To obtain the expression of the physical self-energy, we perform the analytical continuation to the real axis: $\mathrm{i}  \omega_{m^{e}} \rightarrow \omega + \mathrm{i} 0^{+}$. Using the identities (\ref{Id-SumMu1}), (\ref{Id-SumMu2}) and (\ref{Id-SumMu3}) in order to calculate the summations over the imaginary chemical potentials, we obtain at $T=0$
\begin{widetext}
\begin{eqnarray}
& &F_{(3,0)\mbox{\scriptsize{a}}} 
= 2  \left(\frac{J_{\mathrm{K}}}{\mathcal{N}_{s}}\right)^{3} \sum_{\bm{k}_{\bm{1}},\bm{k}_{\bm{2}}} \delta \big(\omega-\omega^{'}\big) 
\frac{1}{(\omega -\varepsilon_{\bm{k}_{\bm{1}}} + \mathrm{i} 0^{+})(\varepsilon_{\bm{k}_{\bm{1}}} - \varepsilon_{\bm{k}_{\bm{2}}}) }
\Bigg[ n_{\mathrm{FD}}(\varepsilon_{\bm{k}_{\bm{2}}})\left(\frac{d_{ \{ \Gamma \} }-1}{\big[d_{ \{ \Gamma \} }\big]^{2} }\right)
- \left(\frac{d_{ \{ \Gamma \} }-1}{\big[d_{ \{ \Gamma \} }\big]^{3} }\right) \Bigg]~,
\label{F30-fin}
\end{eqnarray}
\end{widetext}
where $\omega^{'}$ is substituted to $\mathrm{i}  \omega_{m^{e'}}$ after continuation to the real axis. We have now everything in hand to write the contribution to the electronic self-energy, at zero-temperature. We put together the spin factor (\ref{Contrib-Spin3}) and the frequency-momentum part
(\ref{F30-fin}), and multiply them by two additional factors: (-1) due to the pseudofermion loop, and ($N_{i}$) the number of impurity spins. We get
\begin{eqnarray}
\mbox{\large{$\Sigma$}}_{(3,0)\mbox{\scriptsize{a}}}^{\sigma, \sigma^{'}} \big(\omega, \omega^{'}\big) &=& (-1) \cdot (N_{i}) 
\cdot  \mbox{Spin}\Big[ \mbox{\large{$\Sigma$}}_{(3,0)\mbox{\scriptsize{a}}}^{\sigma, \sigma^{'}} \Big] \nonumber \\
& & \times \ F_{(3,0)\mbox{\scriptsize{a}}}~.
\label{SelfEn3ord-def}
\end{eqnarray}
In the thermodynamic limit we adopt the standard cut-off scheme \cite{Anderson-PMS,Hewson-book} in energy for the conduction electron band: $-D \leq \varepsilon_{\bm{k}} \leq D$, and perform the following substitution
\begin{eqnarray}
\frac{\displaystyle 1}{\displaystyle \mathcal{N}_{s}} \sum\limits_{\bm{k}} \rightarrow \rho \int_{-D}^{+D} d \varepsilon, \nonumber
\end{eqnarray}
where $\rho$ is the density of states per spin and per channel, taken as a constant. We obtain finally 
\begin{widetext}
\begin{eqnarray}
\label{SelfEn3ord-fin} 
& &\mbox{\large{$\Sigma$}}_{(3,0)\mbox{\scriptsize{a}}}^{\sigma, \sigma^{'}} \big(\omega, \omega^{'}\big)  \\
= &-& \Bigg( \Big[\hspace{2mm} (2S)^{2} \big(2-3N+N^{2} \big) +  q^{2} \big(2+3N+N^{2} \big) + 2S \cdot q  \big(4 - N^{2} \big)
+ 2S \big(-4+3N -4N^{2} + 2N^{3}\big) \nonumber \\
& & \hspace{6mm} + q  \big(-4-3N -4N^{2} - 2N^{3}\big) 
+ \big(2+4N^{2}+N^{4}\big) \Big] \frac{(N+2S-1)!}{(2S)! (q-1)! (N-q)!} \cdot \frac{2S}{4 N^{3}} \Bigg) \nonumber \\
&\times& \big( J_{\mathrm{K}} \big)^{3} \cdot n_{i} \cdot \delta_{\sigma, \sigma^{'}} \cdot \delta \big(\omega-\omega^{'}\big) 
\nonumber \\
&\times& 
 \rho^{2} \int_{-D}^{+D} d \varepsilon_{1} \int_{-D}^{+D} d \varepsilon_{2} \
\frac{1}{(\omega -\varepsilon_{1} + \mathrm{i} 0^{+})(\varepsilon_{1} - \varepsilon_{2}) }
\Bigg[ n_{\mathrm{FD}}(\varepsilon_{2})\left(\frac{d_{ \{ \Gamma \} }-1}{\big[d_{ \{ \Gamma \} }\big]^{2} }\right)
- \left(\frac{d_{ \{ \Gamma \} }-1}{\big[d_{ \{ \Gamma \} }\big]^{3} }\right) \Bigg], \hspace{5mm} \nonumber 
\end{eqnarray}
\end{widetext}
where $n_{i} = N_{i} / \mathcal{N}_{s}$ is the impurity density.




\begin{thebibliography}{99}

\bibitem{BlandNoz-1980}
P. Nozi\`eres, and A. Blandin, J. Phys. (Paris) \textbf{41}, 193 (1980).

\bibitem{Noz-1985}
P. Nozi\`eres, Ann. Phys. Fr. \textbf{10}, 19 (1985).

\bibitem{Zawa-1980}
A. Zawadowski, Phys. Rev. Lett. \textbf{45}, 211 (1980).

\bibitem{Zawa-1982}
A. Zawadowski, and K. Vlad\'ar, Phys. Rev. B \textbf{28}, 1564 (1983);
\textit{ibid.} \textbf{28}, 1582 (1983); \textit{ibid.} \textbf{28}, 1596 (1983).

\bibitem{ZarVlad-1996}
G. Zar\'and, and K. Vlad\'ar, Phys. Rev. Lett. \textbf{76}, 2133 (1996).

\bibitem{Zarand-1996}
G. Zar\'and, Phys. Rev. Lett. \textbf{77}, 3609 (1996).

\bibitem{CoqSchri}
B. Coqblin, and J.R. Schrieffer, Phys. Rev. \textbf{185}, 847 (1969).

\bibitem{Furu-1995}
A. Furusaki, and K.A. Matveev, Phys. Rev. B \textbf{52}, 16676 (1995).

\bibitem{jaz}
A. Jerez, N. Andrei, and G. Zar\'and, Phys. Rev. B \textbf{58}, 3814 (1998).

\bibitem{Sachdev-book}
S. Sachdev, \textit{Quantum Phase Transitions}
 (Cambridge University Press, Cambridge, 1999).
 
\bibitem{Lohn-1994}
H.v. L\"ohneysen, T. Pietrus, G. Portisch, H.G. Schlager, A. Schr\"oder, M. Sieck, and T. Trappmann,
Phys. Rev. Lett. \textbf{72}, 3262 (1994).

\bibitem{NozPin}
D. Pines, and P. Nozi\`eres, \textit{The Theory of Quantum Liquids, Volume I - Normal Fermi Liquids} 
(Addison-Wesley Publishing Co., Reading, Massachusetts, 1989).

\bibitem{mathur98}
N.D. Mathur, F.M. Grosche, S.R. Julian, I.R. Walker, D.M. Freye, R.K.W. Haselwimmer, and G.G. Lonzarich,
Nature \textbf{394}, 39 (1998).

\bibitem{trovarelli00}
O. Trovarelli, C. Geibel, S. Mederle, C. Langhammer, F.M. Grosche, P. Gegenwart, M. Lang, G. Sparn, and F. Steglich,
Phys. Rev. Lett. \textbf{85}, 626 (2000).

\bibitem{Stewart-1994}
G.R. Stewart, Rev. Mod. Phys. \textbf{73}, 797 (2001).

\bibitem{Noz-1980}
D.M. Cragg, P. Lloyd, and P. Nozi\`eres,
J. Phys. C \textbf{13}, 803 (1980).

\bibitem{Kusu-1996}
H. Kusunose, K. Miyake, Y. Shimizu, and O. Sakai, Phys. Rev. Lett. \textbf{76}, 271 (1996).

\bibitem{Andrei-1984}
N. Andrei, and C. Destri, Phys. Rev. Lett. \textbf{52}, 364 (1984).

\bibitem{TsveWi-1984}
A.M. Tsvelick, and P.B. Wiegmann, Z. Phys. B \textbf{54}, 201 (1984).

\bibitem{Frad-1990}
E. Fradkin, C. von Reichenbach, and F.A. Schaposnik, 
Nucl. Phys. B \textbf{340}, 692 (1990).

\bibitem{Tsve-1990}
A.M. Tsvelick, J. Phys.: Cond. Matt. \textbf{2}, 2833 (1990).

\bibitem{AfLu-PRL1991}
I. Affleck, and A.W.W. Ludwig, Phys. Rev. Lett. \textbf{67}, 161 (1991).

\bibitem{LuAf-PRL1991}
A.W.W. Ludwig, and I. Affleck, Phys. Rev. Lett. \textbf{67}, 3160 (1991).

\bibitem{Gan-PRL}
J. Gan, N. Andrei, and P. Coleman, Phys. Rev. Lett. \textbf{70}, 686 (1993).

\bibitem{Gan-thesis}
J. Gan, J. Phys.: Cond. Matt. \textbf{6}, 4547 (1994);
J. Gan, \textit{Ph.D Thesis}, Rutgers University (1992).

\bibitem{Hewson-book}
A.C. Hewson,  \textit{The Kondo Problem to Heavy Fermions} (Cambridge University Press, 
Cambridge, 1993).

\bibitem{CoxZaw}
D.L. Cox, and A. Zawadowski, Adv. Phys. \textbf{47}, 599 (1998).

\bibitem{schroder98}
A. Schr\"oder, G. Aeppli, E. Bucher, R. Ramazashvili, and P. Coleman
Phys. Rev. Lett. \textbf{80}, 5623 (1998).

\bibitem{schroder00}
A. Schr\"oder, G. Aeppli, R. Coldea, M. Adams, O. Stockert, H.v. L\"ohneysen, 
E. Bucher, R. Ramazashvili, and P. Coleman, Nature \textbf{407}, 351 (2000).

\bibitem{hertz76}
J.A. Hertz, Phys. Rev. B \textbf{14}, 1165 (1976).

\bibitem{millis93}
A.J. Millis, Phys. Rev. B \textbf{48}, 7183 (1993).

\bibitem{continentino93}
M.A. Continentino, Phys. Rev. B \textbf{47}, 11581 (1993).

\bibitem{moriya95}
T. Moriya, and T. Takimoto, J. Phys. Soc. Jpn \textbf{64}, 960 (1995).

\bibitem{lp00}
M. Lavagna, and C. P\'epin, Phys. Rev. B \textbf{62}, 6450 (2000).

\bibitem{si01}
Q. Si, S. Rabello,  K. Ingersent, and J.L. Smith, Nature \textbf{413}, 804 (2001).

\bibitem{CPT-PRB}
P. Coleman, C. P\'epin, and  A.M. Tsvelik, Phys. Rev. B \textbf{62}, 3852 (2000).

\bibitem{Anderson-PMS}
P.W. Anderson, J. Phys. C  \textbf{3}, 2439 (1970).

\bibitem{Wilson-1974}
K.G. Wilson, in \textit{Nobel Symposia Vol. 24 (1973) - Collective properties of physical systems}
p. 68 (Academic Press, New York, 1974).

\bibitem{Wilson-1975}
K.G. Wilson, Rev. Mod. Phys. \textbf{47}, 773 (1975).

\bibitem{Nozieres-1976}
P. Nozi\`eres, J. Phys. (Paris) \textbf{37}, C1-271 (1976).

\bibitem{Noz-LocFL1974}
P. Nozi\`eres, J. Low Temp. Phys. \textbf{17}, 31 (1974).

\bibitem{Noz-LocFL1975}
P. Nozi\`eres, in \textit{Proc. 14th Int. Conf. on Low temperature Physics} p. 
339, edited by M. Krusius and M. Vuorio (North-Holland, Amsterdam, 1975).

\bibitem{Hewson-1980}
D.M. Newns, and A.C. Hewson, J. Phys. F  \textbf{10}, 2429 (1980).

\bibitem{JLB-Kondo03}
A. Jerez, M. Lavagna, and D. Bensimon, Phys. Rev. B \textbf{68}, 094410 (2003).

\bibitem{JLB-Kondo03Short}
M. Lavagna, A. Jerez, and D. Bensimon, in \textit{Concepts in Electron Correlation} p. 
219, edited by A. C. Hewson and V. Zlati\'c (Kluwer Academic Publisher, Dordrecht, 2003).

\bibitem{CPT-NPB}
P. Coleman, C. P\'epin, and  A.M. Tsvelik, Nucl. Phys. B \textbf{586}, 641 (2000).

\bibitem{ReadNew-JPC1983}
N. Read, and D.M. Newns, J. Phys. C \textbf{16}, 3273 (1983).

\bibitem{Gaudin-1960}
M. Gaudin, Nucl. Phys. \textbf{15}, 89 (1960).

\bibitem{PopovFed}
V.N. Popov, and S.A. Fedotov, Sov. Phys. JETP \textbf{67}, 535 (1988).

\bibitem{FabriZar-1996}
M. Fabrizio, and G. Zar\'and, Phys. Rev. B \textbf{54}, 10008 (1996).

\bibitem{Shaw-1998}
M. Shaw, X.-w. Zhang, and D. Jin, Phys. Rev. B \textbf{57}, 8381 (1998).

\bibitem{AffLud-NPB1991}
I. Affleck, and A.W.W. Ludwig, Nucl. Phys. B \textbf{360}, 641 (1991).

\bibitem{AffLud-1994}
A.W.W. Ludwig, and I. Affleck, Nucl. Phys. B \textbf{428}, 545 (1994).

\bibitem{Natan-1998}
P. Zinn-Justin, and N. Andrei, Nucl. Phys. B \textbf{528}, 648 (1998).

\bibitem{CoRu}
D.L. Cox, and A.E. Ruckenstein, Phys. Rev. Lett. \textbf{71}, 1613 (1993).

\bibitem{pgks}
 O. Parcollet, A. Georges, G. Kotliar, and A. Sengupta,
Phys. Rev. B \textbf{58}, 3794 (1998).

\bibitem{Kondo-1964}
J. Kondo, Prog. Theor. Phys. \textbf{32}, 37 (1964).

\bibitem{Abri1965}
A.A. Abrikosov, Physics \textbf{2}, 5 (1965); \textit{ibid.} \textbf{2}, 61 (1965).

\bibitem{Licht1970}
D.B. Lichtenberg, \textit{Unitary Symmetry and Elementary Particles} (Academic Press,
New York, 1978).

\bibitem{GreiMull}
W. Greiner, and B. M\"uller, \textit{Quantum Mechanics. Symmetries} (Springer-Verlag, Berlin, 2001).

\bibitem{NegOr}
J.W. Negele, and H. Orland, \textit{Quantum Many-Particle Systems}
(Advanced Book Classics, Perseus Books, Reading, Massachusetts, 1998).

\bibitem{Popov-book}
V.N. Popov, \textit{Functional integrals and collective excitations} (Cambridge University Press, Cambridge, 1987).

\bibitem{Mahan} 
G. Mahan, \textit{Many-Particle Physics, second edition} (Plenum Press, New York, 1990).

\bibitem{Mattuck} 
R.D. Mattuck, \textit{A guide to Feynman diagrams in the many-body problem, second edition}
(Dover Publications, New York, 1992).

\bibitem{Silv-Du}
S.D. Silverstein, and C. B. Duke, Phys. Rev. \textbf{161}, 456 (1967).

\bibitem{ItzikZub}
C. Itzykson, and J.B. Zuber, \textit{Quantum Field Theory} (McGraw-Hill, New York, 1980).

\bibitem{LeBellac}
M. Le Bellac, \textit{Quantum and statistical field theory} (Oxford University Press, Oxford, 1991).

\bibitem{Callan}
C.G. Callan, Phys. Rev. D \textbf{2}, 1541 (1970).

\bibitem{Symanzik}
K. Symanzik, Comm. Math. Phys. \textbf{18}, 227 (1970).

\bibitem{AbriMig}
A.A. Abrikosov, and A. A. Migdal, J. Low Temp. Phys. \textbf{3}, 519 (1970).

\bibitem{Loram-1970}
J.W. Loram, T.E. Whall, and P.J. Ford, Phys. Rev. B \textbf{2}, 857 (1970).

\bibitem{Andrei-RMP1983}
N. Andrei, K. Furuya, and J.H. Lowenstein, Rev. Mod. Phys. \textbf{55}, 331 (1983).

\bibitem{Doniach-1977}
S. Doniach, Physica B \textbf{91}, 231 (1977).

\bibitem{Natan-PCom}
N. Andrei (private communication).


 
\bibitem{GellMann-1962}
M. Gell-Mann,  Phys. Rev. \textbf{125}, 1067 (1962).
 
\bibitem{dfTENS-1967}
L.M. Kaplan, and M. Resnikoff, J. Math. Phys. \textbf{8}, 2194 (1967).

\bibitem{dfTENS-1998}
 J.A. de Azc\'arraga, A.J. Macfarlane, A.J. Mountain, and J.C. P\'erez Bueno,
Nucl. Phys. B \textbf{510}, 657 (1998).

\bibitem{IVerona}
I.V. Schensted, \textit{A course on the application of Group theory to Quantum Mechanics}
(NEO Press, Maine, 1976).

\bibitem{Okubo-1977}
S. Okubo, J. Math. Phys. \textbf{18}, 2382 (1977).

\bibitem{Okubo-1982}
S. Okubo, J. Math. Phys. \textbf{23}, 8 (1982).

\bibitem{Schell-1999}
 T. van Ritbergen, A.N. Schellekens, and J.A.M. Vermaseren,
Int. J. Mod. Phys. A \textbf{14}, 41 (1999).

\bibitem{Opper1994}
O. Veits, R. Oppermann, M. Binderberger, and J. Stein,
J. Phys. I France \textbf{4}, 493 (1994).

\bibitem{Kiselev2001}
M.N. Kiselev, H. Feldmann, and R. Oppermann, Eur. Phys. J. B \textbf{22}, 53 (2001).

\bibitem{TejiOgu1995}
S. Tejima, and A. Oguchi, J. Phys. Soc. Jpn. \textbf{64}, 4923 (1995).

\bibitem{BouiKis1999}
F. Bouis, and M.N. Kiselev, Physica B \textbf{259-261}, 195 (1999).

\bibitem{KisOpp-JETP2000}
M.N. Kiselev, and R. Oppermann, JETP Lett. \textbf{71}, 250 (2000).

\bibitem{KisOpp-PRL2000}
M.N. Kiselev, and R. Oppermann,  Phys. Rev. Lett. \textbf{85}, 5631 (2000).

\bibitem{Tsvelik-book1}
A.M. Tsvelik, \textit{Quantum field theory in condensed matter physics} (Cambridge University Press, Cambridge, 1995).

\bibitem{Okubo-1977-2}
S. Okubo, Phys. Rev. D \textbf{16}, 3528 (1977).

\bibitem{OkuPat-1983}
S. Okubo, and J. Patera,  J. Math. Phys. \textbf{24}, 2722 (1983).

\end{thebibliography}
\end{document}